\input harvmac
\input amssym
\input epsf

\lref\CordovaUOB{
  C.~Cordova, D.~S.~Freed, H.~T.~Lam and N.~Seiberg,
  ``Anomalies in the Space of Coupling Constants and Their Dynamical Applications II,''
[arXiv:1905.13361 [hep-th]].
}

\lref\BanksFI{
  T.~Banks and E.~Rabinovici,
  ``Finite Temperature Behavior of the Lattice Abelian Higgs Model,''
Nucl.\ Phys.\ B {\bf 160}, 349 (1979).
}
\lref\FradkinDV{
  E.~H.~Fradkin and S.~H.~Shenker,
  ``Phase Diagrams of Lattice Gauge Theories with Higgs Fields,''
Phys.\ Rev.\ D {\bf 19}, 3682 (1979).
}

\lref\Segal{
G.~Segal, ``The definition of conformal field theory,'' unpublished preprint from late 1980s,
 reprinted in {\sl Topology, geometry and quantum field theory},
    London Math. Soc. Lecture Note  {\bf 308}
     421--577, 2004.
}
\lref\BelovHT{
  D.~Belov and G.~W.~Moore,
  ``Conformal blocks for $AdS_5$ singletons,''
[hep-th/0412167].
}

\lref\WittenAT{
  E.~Witten,
  ``Geometric Langlands From Six Dimensions,''
[arXiv:0905.2720 [hep-th]].
}

\lref\FreedBS{
  D.~S.~Freed and C.~Teleman,
  ``Relative quantum field theory,''
[arXiv:1212.1692 [hep-th]].
}
\lref\MooreJV{
  G.~W.~Moore,
  ``Anomalies, Gauss laws, and Page charges in M-theory,''
Comptes Rendus Physique {\bf 6}, 251 (2005).
[hep-th/0409158].
}

\lref\DixonJC{
  L.~J.~Dixon, J.~A.~Harvey, C.~Vafa and E.~Witten,
  ``Strings on Orbifolds. 2.,''
Nucl.\ Phys.\ B {\bf 274}, 285 (1986).
}

\lref\SeibergPQ{
  N.~Seiberg,
  ``Electric--magnetic duality in supersymmetric nonAbelian gauge theories,''
Nucl.\ Phys.\ B {\bf 435}, 129 (1995).
[hep-th/9411149].
}

\lref\BanksZN{
  T.~Banks and N.~Seiberg,
  ``Symmetries and Strings in Field Theory and Gravity,''
Phys.\ Rev.\ D {\bf 83}, 084019 (2011).
[arXiv:1011.5120 [hep-th]].
}

\lref\SeibergDR{
  N.~Seiberg and W.~Taylor,
  ``Charge Lattices and Consistency of 6D Supergravity,''
JHEP {\bf 1106}, 001 (2011).
[arXiv:1103.0019 [hep-th]].
}

\lref\GirardelloGF{
  L.~Girardello, A.~Giveon, M.~Porrati and A.~Zaffaroni,
  ``S duality in N=4 Yang-Mills theories with general gauge groups,''
Nucl.\ Phys.\ B {\bf 448}, 127 (1995).
[hep-th/9502057].
}

\lref\WuBV{
  S.~Wu,
  ``S-duality in Vafa-Witten theory for non-simply laced gauge groups,''
JHEP {\bf 0805}, 009 (2008).
[arXiv:0802.2047 [hep-th]].
}

\lref\StrasslerFE{
  M.~J.~Strassler,
  ``Duality, phases, spinors and monopoles in $SO(N)$ and $spin(N)$ gauge theories,''
JHEP {\bf 9809}, 017 (1998).
[hep-th/9709081].
}

\lref\ARSW{O.~Aharony, S.~S.~Razamat, N.~Seiberg and B.~Willett,
  ``3d dualities from 4d dualities,''
JHEP {\bf 1307}, 149 (2013).
[arXiv:1305.3924 [hep-th]]. O.~Aharony, S.~S.~Razamat, N.~Seiberg and B.~Willett,``3d dualities from 4d dualities for orthogonal groups,''
[arXiv:1307.0511 [hep-th]].}

\lref\IntriligatorER{
  K.~A.~Intriligator and N.~Seiberg,
  ``Phases of $\CN=1$ supersymmetric gauge theories and electric-magnetic triality,''
In {\sl Los Angeles 1995, Future perspectives in string theory} 270-282.
[hep-th/9506084].
}
\lref\IntriligatorID{
  K.~A.~Intriligator and N.~Seiberg,
  ``Duality, monopoles, dyons, confinement and oblique confinement in supersymmetric $SO(N_c)$ gauge theories,''
Nucl.\ Phys.\ B {\bf 444}, 125 (1995).
[hep-th/9503179].
}

\lref\mooretalks{
G.~Moore, ``Overview of the theory of self-dual fields'' and ``The RR charge of an orientifold'', which are available through http://www.physics.rutgers.edu/\~{}gmoore/ .
}
\lref\AharonyQU{
  O.~Aharony and E.~Witten,
  ``Anti-de Sitter space and the center of the gauge group,''
JHEP {\bf 9811}, 018 (1998).
[hep-th/9807205].
}

\lref\IntriligatorAU{
  K.~A.~Intriligator and N.~Seiberg,
  ``Lectures on supersymmetric gauge theories and electric--magnetic duality,''
Nucl.\ Phys.\ Proc.\ Suppl.\  {\bf 45BC}, 1 (1996).
[hep-th/9509066].
}

\lref\CachazoZK{
  F.~Cachazo, N.~Seiberg and E.~Witten,
  ``Phases of $\CN=1$ supersymmetric gauge theories and matrices,''
JHEP {\bf 0302}, 042 (2003).
[hep-th/0301006].
}

\lref\DoldWhitney{
  A.~Dold and H.~Whitney,
  ``Classification of oriented sphere bundles over a 4-complex,''
  Ann.~of Math.\ 2nd Ser. {\bf 69}, 667--677 (1959)
}
\lref\Thomas{
  E.~Thomas,
  ``On the Cohomology of the Real Grassmann Complexes and the Characteristic Classes of $n$-Plane Bundles,''
  Trans.~Amer.~Math.~Soc. {\bf 96}, 67--89 (1960)
}
\lref\Borel{
  A.~Borel,
  ``Sur L'Homologie et la Cohomologie des Groupes de Lie Compacts Connexes,''
  Jour.~Amer.~Math. {\bf 76}, 273--342 (1954)
}
\lref\KMS{
A.~Kono, M.~Mimura, and N.~Shimada,
``Cohomology of classifying spaces of certain associative $H$-spaces,''
J.~Math.~Kyoto.~U. {\bf 15} 607--617 (1975)
}
\lref\KM{
A.~Kono, and M.~Mimura,
``On the cohomology mod 2 of the classifying space of $Ad E_7$,''
J.~Math.~Kyoto.~U. {\bf 18} 535--541 (1978)
}
\lref\BrowderThomas{
W.~Browder and E.~Thomas,
``Axioms for the generalized Pontryagin cohomology operations,''
Quart.~J.~Math.~Oxford, 2nd ser., {\bf 13} 55--60 (1962)
}

\lref\GaiottoBE{
  D.~Gaiotto, G.~W.~Moore and A.~Neitzke,
  ``Framed BPS States,''
[arXiv:1006.0146 [hep-th]].
}
\lref\WittenNV{
  E.~Witten,
  ``Supersymmetric index in four-dimensional gauge theories,''
Adv.\ Theor.\ Math.\ Phys.\  {\bf 5}, 841 (2002).
[hep-th/0006010].
}

\lref\KapustinPY{
  A.~Kapustin,
  ``Wilson-'t Hooft operators in four-dimensional gauge theories and S-duality,''
Phys.\ Rev.\ D {\bf 74}, 025005 (2006).
[hep-th/0501015].
}

\lref\SeibergRS{
  N.~Seiberg and E.~Witten,
  ``Electric-magnetic duality, monopole condensation, and confinement in $\CN=2$ supersymmetric Yang-Mills theory,''
Nucl.\ Phys.\ B {\bf 426}, 19 (1994), [Erratum-ibid.\ B {\bf 430}, 485 (1994)].
[hep-th/9407087].
}

\lref\GoddardQE{
  P.~Goddard, J.~Nuyts and D.~I.~Olive,
  ``Gauge Theories and Magnetic Charge,''
Nucl.\ Phys.\ B {\bf 125}, 1 (1977).
}

\lref\GukovJK{
  S.~Gukov and E.~Witten,
  ``Gauge Theory, Ramification, And The Geometric Langlands Program,''
[hep-th/0612073].
}

\lref\GukovSN{
  S.~Gukov and E.~Witten,
  ``Rigid Surface Operators,''
Adv.\ Theor.\ Math.\ Phys.\  {\bf 14} (2010).
[arXiv:0804.1561 [hep-th]].
}

\lref\WittenXY{
  E.~Witten,
  ``Baryons and branes in anti-de Sitter space,''
JHEP {\bf 9807}, 006 (1998).
[hep-th/9805112].
}

\lref\HananyFQ{
  A.~Hanany and B.~Kol,
  ``On orientifolds, discrete torsion, branes and M theory,''
JHEP {\bf 0006}, 013 (2000).
[hep-th/0003025].
}

\lref\GanorNF{
  O.~J.~Ganor,
  ``Six-dimensional tensionless strings in the large N limit,''
Nucl.\ Phys.\ B {\bf 489}, 95 (1997).
[hep-th/9605201].
}

\lref\HenningsonHP{
  M.~Henningson,
  ``Wilson-'t Hooft operators and the theta angle,''
JHEP {\bf 0605}, 065 (2006).
[hep-th/0603188].
}

\lref\IntriligatorNE{
  K.~A.~Intriligator and P.~Pouliot,
  ``Exact superpotentials, quantum vacua and duality in supersymmetric $Sp(N_c)$ gauge theories,''
Phys.\ Lett.\ B {\bf 353}, 471 (1995).
[hep-th/9505006].
}

\lref\BanksZN{
  T.~Banks and N.~Seiberg,
  ``Symmetries and Strings in Field Theory and Gravity,''
Phys.\ Rev.\ D {\bf 83}, 084019 (2011).
[arXiv:1011.5120 [hep-th]].
}

\lref\CaldararuTC{
  A.~Caldararu, J.~Distler, S.~Hellerman, T.~Pantev and E.~Sharpe,
  ``Non-birational twisted derived equivalences in abelian GLSMs,''
  arXiv:0709.3855 [hep-th].
}

\lref\PantevRH{
  T.~Pantev and E.~Sharpe,
  ``Notes on gauging noneffective group actions,''
  arXiv:hep-th/0502027.
}

\lref\PantevZS{
  T.~Pantev and E.~Sharpe,
  ``GLSM's for gerbes (and other toric stacks),''
  Adv.\ Theor.\ Math.\ Phys.\  {\bf 10}, 77 (2006)
  [arXiv:hep-th/0502053].
}

\lref\WittenWY{
  E.~Witten,
  ``AdS / CFT correspondence and topological field theory,''
JHEP {\bf 9812}, 012 (1998).
[hep-th/9812012].
}

\lref\SeibergQD{
  N.~Seiberg,
  ``Modifying the Sum Over Topological Sectors and Constraints on Supergravity,''
JHEP {\bf 1007}, 070 (2010).
[arXiv:1005.0002 [hep-th]].
}

\lref\MaldacenaRE{
  J.~M.~Maldacena,
  ``The Large N limit of superconformal field theories and supergravity,''
Adv.\ Theor.\ Math.\ Phys.\  {\bf 2}, 231 (1998).
[hep-th/9711200].
}

\lref\MaldacenaSS{
  J.~M.~Maldacena, G.~W.~Moore and N.~Seiberg,
  ``D-brane charges in five-brane backgrounds,''
JHEP {\bf 0110}, 005 (2001).
[hep-th/0108152].
}

\lref\ShifmanYH{
  M.~Shifman and A.~Yung,
  ``Confronting Seiberg's Duality with $r$ Duality in ${\cal N}=1$ Supersymmetric QCD,''
Phys.\ Rev.\ D {\bf 86}, 065003 (2012).
[arXiv:1204.4164 [hep-th]].
}

\lref\KapustinWM{
  A.~Kapustin and N.~Saulina,
  ``The Algebra of Wilson-'t Hooft operators,''
Nucl.\ Phys.\ B {\bf 814}, 327 (2009).
[arXiv:0710.2097 [hep-th]].
}

\lref\ShifmanEWA{
  M.~Shifman and A.~Yung,
  ``Detailing N=1 Seiberg's Duality through the Seiberg-Witten Solution of N=2,''
[arXiv:1304.0822 [hep-th]].
}

\lref\KonishiTX{
  K.~Konishi and Y.~Ookouchi,
  ``On Confinement Index,''
Nucl.\ Phys.\ B {\bf 827}, 59 (2010).
[arXiv:0909.3781 [hep-th]].
}

\lref\VafaTF{
  C.~Vafa and E.~Witten,
  ``A strong coupling test of S duality,''
Nucl.\ Phys.\ B {\bf 431}, 3 (1994).
[hep-th/9408074].
}

\lref\VafaWX{
  C.~Vafa,
  ``Modular Invariance and Discrete Torsion on Orbifolds,''
Nucl.\ Phys.\ B {\bf 273}, 592 (1986).
}

\lref\VafaRV{
  C.~Vafa and E.~Witten,
  ``On orbifolds with discrete torsion,''
J.\ Geom.\ Phys.\  {\bf 15}, 189 (1995).
[hep-th/9409188].
}

\lref\ArgyresHC{
  P.~C.~Argyres, M.~R.~Plesser and A.~D.~Shapere,
  ``N=2 moduli spaces and N=1 dualities for $SO(n_c)$ and $USp(2n_c)$ superQCD,''
Nucl.\ Phys.\ B {\bf 483}, 172 (1997).
[hep-th/9608129].
}

\lref\ArgyresEH{
  P.~C.~Argyres, M.~R.~Plesser and N.~Seiberg,
  ``The Moduli space of vacua of N=2 SUSY QCD and duality in N=1 SUSY QCD,''
Nucl.\ Phys.\ B {\bf 471}, 159 (1996).
[hep-th/9603042].
}

\lref\ArgyresQR{
  P.~C.~Argyres, A.~Kapustin and N.~Seiberg,
  ``On S-duality for non-simply-laced gauge groups,''
JHEP {\bf 0606}, 043 (2006).
[hep-th/0603048].
}

\lref\DrukkerTZ{
  N.~Drukker, D.~R.~Morrison and T.~Okuda,
  ``Loop operators and S-duality from curves on Riemann surfaces,''
JHEP {\bf 0909}, 031 (2009).
[arXiv:0907.2593 [hep-th]].
}

\lref\AharonyTI{
  O.~Aharony, S.~S.~Gubser, J.~M.~Maldacena, H.~Ooguri and Y.~Oz,
  ``Large N field theories, string theory and gravity,''
Phys.\ Rept.\  {\bf 323}, 183 (2000).
[hep-th/9905111].
}

\lref\DijkgraafPZ{
  R.~Dijkgraaf and E.~Witten,
  ``Topological Gauge Theories and Group Cohomology,''
Commun.\ Math.\ Phys.\  {\bf 129}, 393 (1990).
}

\lref\SeibergNZ{
  N.~Seiberg and E.~Witten,
  ``Gauge dynamics and compactification to three-dimensions,''
In {\sl Saclay 1996, The mathematical beauty of physics} 333-366.
[hep-th/9607163].
}

\lref\EvansHI{
  N.~J.~Evans, S.~D.~H.~Hsu and M.~Schwetz,
  ``Phase transitions in softly broken N=2 SQCD at nonzero theta angle,''
Nucl.\ Phys.\ B {\bf 484}, 124 (1997).
[hep-th/9608135].
}

\lref\KonishiIZ{
  K.~Konishi,
  ``Confinement, supersymmetry breaking and theta parameter dependence in the Seiberg-Witten model,''
Phys.\ Lett.\ B {\bf 392}, 101 (1997).
[hep-th/9609021].
}

\lref\tHooftHT{
  G.~'t Hooft,
  ``Topology of the Gauge Condition and New Confinement Phases in Nonabelian Gauge Theories,''
Nucl.\ Phys.\ B {\bf 190}, 455 (1981).
}


\def\g{{\bf g}}

\def\bC{{\bf C}}
\def\bH{{\bf H}}
\def\bW{{\bf W}}


\def\bb{
\font\tenmsb=msbm10
\font\sevenmsb=msbm7
\font\fivemsb=msbm5
\textfont1=\tenmsb
\scriptfont1=\sevenmsb
\scriptscriptfont1=\fivemsb
}



\def\vev#1{\left\langle #1\right\rangle}


\def\tilde{\widetilde}

\def\hat{\widehat}

\def\bar{\overline}
\def\b{\bar}
\def\bsq#1{{{\b{#1}}^{\lower 2.5pt\hbox{$\scriptstyle 2$}}}}
\def\bexp#1#2{{{\b{#1}}^{\lower 2.5pt\hbox{$\scriptstyle #2$}}}}
\def\dotexp#1#2{{{#1}^{\lower 2.5pt\hbox{$\scriptstyle #2$}}}}


\def\rt2{\sqrt{2}}

\def\Re{\mathop{\rm Re}}

\def\mod{{\rm mod}}
\def\det{\mathop{\rm det}}



\def\CN{{\cal N}}

\def\CP{{\cal P}}


\def\1{{\ds 1}}
\def\R{\hbox{$\bb R$}}

\def\Z{\hbox{$\bb Z$}}

\def\S{\hbox{$\bb S$}}
\def\T{\hbox{$\bb T$}}


\noblackbox

\def\unit{\relax{\rm 1\kern-.26em I}}
\def\nada{\relax{\rm 0\kern-.30em l}}
\def\tilde{\widetilde}

\def\mod{{\rm mod}}
\def\CP{{\cal P}}
\noblackbox
\def\IL{\relax{\rm I\kern-.18em L}}
\def\IH{\relax{\rm I\kern-.18em H}}
\def\IR{\relax{\rm I\kern-.18em R}}
\def\IC{\relax\hbox{$\inbar\kern-.3em{\rm C}$}}
\def\IZ{\relax\ifmmode\mathchoice
{\hbox{\cmss Z\kern-.4em Z}}{\hbox{\cmss Z\kern-.4em Z}} {\lower.9pt\hbox{\cmsss Z\kern-.4em Z}}
{\lower1.2pt\hbox{\cmsss Z\kern-.4em Z}}\else{\cmss Z\kern-.4em Z}\fi}

\def\CN {{\cal N}}

\def\partialslash{\not{\hbox{\kern-2pt $\partial$}}}
\def\CP {{\cal P }}


\def\CN {{\cal N}}

\def\CP {{\cal P }}

\font\manual=manfnt \def\dbend{\lower3.5pt\hbox{\manual\char127}}

\def\IZ{\relax\ifmmode\mathchoice
{\hbox{\cmss Z\kern-.4em Z}}{\hbox{\cmss Z\kern-.4em Z}} {\lower.9pt\hbox{\cmsss Z\kern-.4em Z}}
{\lower1.2pt\hbox{\cmsss Z\kern-.4em Z}}\else{\cmss Z\kern-.4em Z}\fi}

\def\bar{\overline}

\def\rt2{\sqrt{2}}
\def\irt2{{1\over\sqrt{2}}}

\def\hat{\widehat}
\def\slashchar#1{\setbox0=\hbox{$#1$}           
   \dimen0=\wd0                                 
   \setbox1=\hbox{/} \dimen1=\wd1               
   \ifdim\dimen0>\dimen1                        
      \rlap{\hbox to \dimen0{\hfil/\hfil}}      
      #1                                        
   \else                                        
      \rlap{\hbox to \dimen1{\hfil$#1$\hfil}}   
      /                                         
   \fi}

\def\gcd{\hbox{gcd}}
\def\ns#1#2#3#4{_{#1#2\atop#3#4}}

\def\figcaption#1#2{\DefWarn#1\xdef#1{Figure~\noexpand\hyperref{}{figure}%
{\the\figno}{\the\figno}}\writedef{#1\leftbracket Figure\noexpand~\xfig#1}%
\medskip\centerline{{\footnotefont\bf Figure~\hyperdef\hypernoname{figure}{\the\figno}{\the\figno}:}  #2 \wrlabeL{#1=#1}}%
\global\advance\figno by1}



\Title {\vbox{\hbox{WIS/03/13-APR-DPPA,\ UT-13-15,\ IPMU13-0081}}}
{\vbox{\centerline{Reading between the lines of}
\vskip7pt
\centerline{four-dimensional gauge theories}} }
 \centerline{Ofer Aharony$^{1,2}$, Nathan Seiberg$^2$ and Yuji Tachikawa$^{3}$}

\bigskip
\centerline{$^{1}$Department of Particle Physics and Astrophysics, Weizmann Institute of Science,}
\centerline{Rehovot 76100, Israel}
\medskip
\centerline{$^{2}$ School of Natural Sciences, Institute for Advanced Study, Princeton, NJ, 08540, USA}
\medskip
\centerline{$^{3}$ Department of Physics, University of Tokyo, 7-3-1 Hongo, Tokyo, 113-0033, Japan and}
\centerline{IPMU, University of Tokyo, Kashiwa, Chiba 277-8583, Japan}
\medskip

\vskip20pt

\noindent
Starting with a choice of a gauge group in four dimensions, there is often freedom in the choice of magnetic  and dyonic line operators.  Different consistent choices of these operators correspond to distinct physical theories, with the same correlation functions of local operators in $\R^4$.  In some cases these choices are permuted by shifting the $\theta$-angle by $2\pi$.  In other cases they are labeled by new discrete $\theta$-like parameters.  Using this understanding we gain new insight into the dynamics of four-dimensional gauge theories and their phases.  The existence of these distinct theories clarifies a number of issues in electric/magnetic dualities of supersymmetric gauge theories, both for the conformal $\CN=4$ theories and for the low-energy dualities of $\CN=1$ theories.

\Date{May 2013}

\newsec{Introduction}

In this paper we study the line operators in four-dimensional gauge theories.  The analysis of a gauge theory starts by choosing a Lie algebra $\g$ and a gauge group $G$.  This determines the allowed Wilson line operators in the theory -- they are in one to one correspondence with the representations of $G$. Below we will analyze the consistency conditions on the magnetic and dyonic line operators. Typically there are several distinct choices for the same $G$.  These different choices correspond to distinct physical theories.

In some cases (like $G=SU(N)/\Z_N$) these different choices can be labeled by extending the range of the $\theta$-angle -- the different theories are permuted by shifting the $\theta$-angle by $2\pi$.  In other cases (like $G=SO(N)$ with $N>4$) the distinct choices are labeled by new discrete $\theta$-like parameters.

The correlation functions of local operators in $\R^4$ depend only on the choice of the Lie algebra $\g$ of the gauge group $G$.  They are independent of the global structure of $G$ and the different choices of line operators.  So naively these subtleties are of no interest for a four-dimensional physicist.  However, we will argue that they have several important consequences.  First, these subtleties affect the correlation functions of line operators in the theory.  Therefore, they affect the phase structure of the theory on $\R^4$. Second, these subtleties become more dramatic when we compactify the theory.  For example, we will see that the choices of $G$ and of these parameters have important consequences even for local dynamics on $\R^3\times \S^1$. In particular, these different theories can have a different number of vacua (and, in supersymmetric theories, different Witten indices) on $\R^3\times \S^1$.  The simple reason for the difference between $\R^4$ and $\R^3\times \S^1$ is that wrapping a line operator around the $\S^1$ leads to a local operator in $\R^3$.
These issues play an important role in the relation between IR dualities of four dimensional gauge theories and those of three dimensional gauge theories \ARSW.

Using this understanding, we revisit a number of issues associated with electric/magnetic duality in four dimensions.  In particular, starting with some $\CN=4$ supersymmetric Yang-Mills (SYM) theory, we can vary its coupling constant and look for various dual weak coupling limits.  We will show that in some situations we uncover distinct theories that have so far been viewed as identical.  Similarly, a number of phenomena involving  the IR duality of $\CN=1$ supersymmetric gauge theories \refs{\SeibergPQ,\IntriligatorID} will be clarified.  Specifically, the duality of $SO(N)$ gauge theories with $N_f$ vectors turns out to be particularly rich and interesting.

In this rather extensive introduction, we will use gauge theories with gauge algebra $su(2)$ as examples to illustrate the main features of our discussions. The following sections contain the generalization of our analysis to other gauge groups.

\subsec{The line operators}

Let us begin by analyzing the line operators.  Consider a four dimensional gauge theory based on the Lie algebra $\g$ and the gauge group $G$. We will only discuss connected gauge groups in this paper. We denote the universal cover of the gauge group $G$ by $\tilde G$, and the center of $\tilde G$ by $\bC$. The gauge group is the quotient $G=\tilde G/\bH$, with $\bH \subset \bC$ a subgroup of the center. We will refer to these groups as electric groups.

For a given gauge algebra $\g$, the choice of the gauge group $G$ determines some of the properties of the gauge theory:
\item{1.} The allowed matter fields must be in representations of $G$; i.e.\ they should be invariant under $\bH$.  Note that the allowed representations are determined by $G$ rather than the other way around.  In particular, if $\bH$ is nontrivial, a theory based on $G$ differs from a theory based on $\tilde G$, even if no matter fields are present.
\item{2.} The Wilson line observables are labeled by representations of $G$; i.e.\ they should be invariant under $\bH$.  Unlike the choice of matter fields, the Wilson line representations include all representations of $G$, regardless of whether dynamical matter fields in these representations are present.  In terms of the weight lattice $\Lambda_w$ of $\g$, the Wilson lines are in one to one correspondence with points in $\Lambda^G_w/\bW$, where $\Lambda^G_w\subset \Lambda_w$ is the sublattice of weights of $G$, and $\bW$ is the Weyl group.
\item{3.} The gauge bundles we should sum over are also affected by the choice of $G$.  The bundles are labeled by certain discrete choices.  If the gauge group is $\tilde G$, the different bundles are labeled by the instanton number $\ell \in \Z$.  If the gauge group is $G$, there are additional distinct bundles -- $G$ bundles that are not $\tilde G$ bundles.  For these, the instanton number $\ell$ can be rational rather than integer.  One parameter that labels distinct theories with the same gauge group $G$ is the $\theta$-angle, which determines how to sum the contributions of the bundles with various values of $\ell$.  Their weights are the phase $e^{i\ell\theta}$.  If the gauge group is $\tilde G$ then the periodicity of $\theta$ is $2\pi$.  Otherwise it can be larger.  Below we will consider additional characteristics of the bundles such as their Stiefel-Whitney classes $w_{2}$ and $w_4$, and we will show that the sum over different bundles can lead to new discrete $\theta$-like parameters, that distinguish between distinct theories with the same gauge group $G$.

 In addition to the Wilson line operators, there are also 't Hooft line operators.
Let us denote by $\g^*$ the Langlands-dual (GNO-dual \GoddardQE) Lie algebra of $\g$,
and by $\tilde G^*$ the simply-connected group with Lie algebra $\g^*$.
Then, 't Hooft lines are labeled by points in
$\Lambda_{mw}/\bW$, where $\Lambda_{mw}$ is the magnetic weight lattice, i.e.~the weight lattice of $\g^*$, which is the dual of the root lattice of $\g$.
More generally, we can have dyonic line operators carrying both electric and magnetic charges. They are labeled by a pair of weights \KapustinPY
\eqn\dyonl{(\lambda_e, \lambda_m) \in {\Lambda_w \times \Lambda_{mw}}}
with the identification
\eqn\ident{(\lambda_e, \lambda_m) \sim (w\lambda_e, w\lambda_m),\qquad w\in \bW}
where $\bW$ is the Weyl group both of $\g$ and $\g^*$, and it acts on the two lattices in \dyonl.  This labeling contains more information than a pair of representations of $\g$ and $\g^*$, which are labeled by $(\Lambda_w/\bW) \times (\Lambda_{mw}/\bW)$  \KapustinPY.

Given the gauge group $G=\tilde G/\bH$,
which line operators labeled by $(\lambda_e, \lambda_m)$ are present in the theory?
First, when the gauge group is $G$
we should include all lines $(\lambda_e,0)$ with $\lambda_e \in \Lambda^G_w/\bW$.
In particular,
the lines $(r_e,0)$, with $r_e$ a root, should be present. They represent the world lines of gauge fields.
Similarly, the purely magnetic lines $(\lambda_e=0,r_m)$ with $r_m$ a root of $\g^*$ must be present.
Furthermore, if $(\lambda_e, \lambda_m)$ and $(\lambda'_e, \lambda'_m)$ are present, then so is $(\lambda_e+\lambda'_e, \lambda_m+ \lambda'_m)$ \foot{This is related to the operator product expansion of line operators, discussed in \KapustinWM.}. Also, if $(\lambda_e, \lambda_m)$ is present, then so is the
orientation-reversed line $(-\lambda_e, -\lambda_m)$.

Therefore, we can consider the charges $(\lambda_e,\lambda_m)$ modulo the root lattice of $\g\times \g^*$. Recalling that the weight lattice modulo the root lattice of the algebra $\g$ is the center $\bC$ of the group $\tilde G$, and that $\tilde G$ and $\tilde G^*$ have the same center, we see that we can organize the line operators into classes labeled by a pair\foot{Throughout our discussion we will carelessly ignore the distinction between elements of an Abelian discrete group and the characters of that group.} $(z_e,z_m) \in \bC\times \bC$. The properties above imply that if one element in a class exists, so do all the elements $(\lambda_e,\lambda_m)$ in that class.  And, the allowed classes should be closed under multiplication and inversion.

As we mentioned above, the choice of a gauge group $G$ implies that the lines $(\lambda_e,0)$ with $\lambda_e \in \Lambda_w^G/\bW$ a weight of the group $G$ should be present. This means that we should include all classes
$(z_e , z_m=0)$, where the $z_e$ are the center charges that are invariant under $\bH$.  A crucial point is that even after this choice is made, there are still distinct theories, labeled by the complete choice of classes of allowed line operators $(z_e,z_m)$ \GaiottoBE.

The allowed choices for which line operators are present are restricted by a version of the Dirac quantization condition \refs{\KapustinPY,\GaiottoBE}.
The correlation function of two lines $\gamma$ and $\gamma'$ should depend only on the positions and representations of $\gamma$ and $\gamma'$. If we hold $\gamma$ fixed, and move the path $\gamma'$ such that it loops around $\gamma$ (along a surface that has nontrivial linking number with $\gamma$) and then comes back to the same position, then the correlation function is multiplied by a phase. For the correlation function to be well-defined, this phase must be equal to one.
This phase depends on the charges $(z_e,z_m)$ and $(z'_e,z'_m)$ of the two lines as follows.
Let us consider the case when $\bC=\Z_{k}$; the case of $\bC=\Z_2\times \Z_2$ will be treated in section 5, and the general case in section 6.3. For $\bC=\Z_k$ the representations of lines are specified by
\eqn\charges{(z_e,z_m)=(n,m)\ \mod\ k~.}
Then, the condition that there is no nontrivial phase as the family of curves $\gamma'$ links $\gamma$ once is given by
\eqn\lcalr{ {n m' - m n'} = 0\, \mod\, k ~.}
This condition guarantees that the correlation functions of line operators are local\foot{Similar conditions are present also for field theories with no known Lagrangian, such as the so-called class S theories, where the line operators still have a similar classification \refs{\DrukkerTZ,\GaiottoBE}. We will not discuss such theories here.}.

\subsec{Example: the line operators for $\g=su(2)$}

As a first example, consider gauge theories based on the Lie algebra $\g= su(2)$, with center $\bC=\Z_2$.
Let us first consider the simply-connected case:
\item{$SU(2)$} Here, the spectrum of operators includes Wilson lines in the fundamental representation $(\lambda_e, \lambda_m)=(1,0)$, or $(z_e,z_m)=(1,0)$. Locality determines the remaining lines to be $(\lambda_e, \lambda_m )$ with $\lambda_e \in \Z$, $\lambda_m \in 2\Z $, so no other nontrivial representations of the center are allowed.

\noindent Let us instead start with the electric gauge group $G=SO(3)$.  The purely electric lines are now $(\lambda_e, \lambda_m=0)$ with $\lambda_e \in 2\Z$.  This set of lines can be completed in two different ways, leading to two distinct theories \GaiottoBE:
\item{$SO(3)_+$} Here the line operators are $(\lambda_e,\lambda_m)$ with $\lambda_e \in 2\Z$,  $\lambda_m \in \Z$. In other words, they have $(z_e,z_m)=(0,0)$ or $(z_e,z_m)=(0,1)$, including the 't Hooft line operator in the fundamental representation of the dual gauge group.
\item{$SO(3)_-$} Here the line operators are $(\lambda_e,\lambda_m)$ with $\lambda_e,\lambda_m \in \Z$ such that $\lambda_e+\lambda_m \in 2\Z $; they have $(z_e,z_m)=(0,0)$ or $(z_e,z_m)=(1,1)$.  In particular, the purely electric line $(1,0)$ and the purely magnetic line $(0,1)$ are not present, but the dyonic line $(1,1)$ is present.

\midinsert
\tabskip1em plus1fill\relax
\halign to\hsize{  \hfil #  \hfil & \hfil # \hfil & \hfil # \hfil \cr
\epsffile{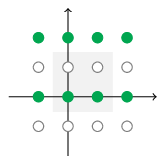} & \epsffile{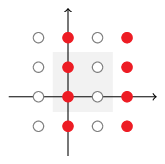} & \epsffile{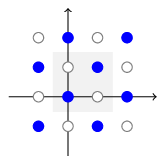} \cr
$SU(2)$ & $SO(3)_+$  & $SO(3)_-$ \cr}
\figcaption\sutwolines{The weights of line operators of gauge theories with $\g=su(2)$.}
\endinsert

\noindent The weights of the available line operators of the three choices, $SU(2)$ and $SO(3)_\pm$, are shown in \sutwolines. There, the horizontal axis is for $\lambda_e$ and the vertical axis is for $\lambda_m$. The shaded regions in the figure give the $\Z_2$ charges.

We presented the two $SO(3)$ theories through their different line operators. Alternatively, they can be described by shifting $\theta$ by $2 \pi$ :
\eqn\sopm{SO(3)_{+}^{\theta}=SO(3)_{-}^{{\theta+2\pi}}~.}
Indeed, the Witten effect shows that under $\theta \to \theta+2\pi$, $(\lambda_e ,\lambda_m) \to (\lambda_e+\lambda_m,\lambda_m)$, which leads to \sopm.
This means that when $G=SO(3)$ the periodicity of $\theta$ is $4\pi$.
This is due to the fact that on spin manifolds\foot{On non-spin manifolds  there can be ``quarter instantons'' and the periodicity of $\theta$ is $8\pi$. }, the instanton number of $SO(3)$ gauge theories is a multiple of $1\over 2$.
Naively, the shift of $\theta$ by $2 \pi$ does not change the local physics.  But since the insertion of the line operators in $\R^4$ creates a nontrivial topology, it allows us to distinguish $\theta$ from $\theta+2\pi$ locally on $\R^4$.
Note that  the insertion of lines in $\R^4$ keeps it a spin manifold, and therefore shifting $\theta$ by $4\pi$ maps the theory to itself, relabeling the line operators.

As we will see amply below, we {\it cannot always} map one choice of line operators to another by a shift of the conventional $\theta$-angle. For example, in section 3 we will see that $SO(N)_+$ and $SO(N)_-$ are not related by a shift of the $\theta$-angle when $N\ge 5$.

\subsec{Analogies with $2d$ orbifolds and other constructions}

The discussion so far is reminiscent of orbifolds in two-dimensional field theories.  There we start with a system with a global discrete symmetry $\Gamma$.  The orbifold is constructed by turning $\Gamma$ into a gauge symmetry.  This has the effect of projecting on the $\Gamma$-invariant states and adding the twisted sector states.  This is similar to our discussion above.  Starting with a gauge theory based on the simply connected group $\tilde G$ we have Wilson lines for all representations of $\g$.  When the gauge group is $G=\tilde G/\bH$ we project on Wilson lines associated with representations that are invariant under $\bH$.  The various magnetic and dyonic lines are similar to the twisted sector states.  The set of allowed operators in the orbifold is restricted by mutual locality. This is analogous to the use of Dirac quantization on the lines. In two dimensions the need to add the twisted sector states follows from modular invariance.
In four dimensions we will also argue that the spectrum of line operators should be complete, including a maximal set of allowed charges\foot{For example, in our $su(2)$ example above, this means that we cannot have a theory with only the $(z_e=0,z_m=0)$ line operators.}.
Finally, it is common in orbifolds that the details of the twisted sector can depend on additional data -- discrete torsion \refs{\VafaWX,\VafaRV}.  This is analogous to our different distinct theories with the same gauge group $G$. We will see in section 6 that the choice of the line operators of four-dimensional gauge theories corresponds to a phase in the path integral that is very similar to the
one that distinguishes theories with discrete torsion. The analogy with orbifolds becomes more complete in sections 6.1 and 6.4. Just as orbifolds correspond to gauging $\Gamma$, we will show that by gauging an appropriate symmetry we can move between our different theories.

Unlike two-dimensional orbifolds, our different theories with the same gauge algebra $\g$ have the same local operators, and they differ only in their line and surface operators.
Similar phenomena were described in \refs{\SeibergQD,\BanksZN} (see also \refs{\PantevZS\PantevRH-\CaldararuTC}).  The phenomena described in these papers have two complementary descriptions.  First, it is a modification of the sum over different bundles.  Second, it is equivalent to coupling a quantum field theory to a discrete gauge symmetry.  The latter symmetry can be an ordinary gauge symmetry whose holonomies are associated with lines, or it can be a higher form symmetry whose holonomies are associated with surfaces or higher dimensional generalizations of them. Our discussion in section 6.4 will make the analogy clearer by presenting our construction in terms of gauging and correspondingly, there will be holonomies on lines and surfaces.

\subsec{The classification of phases of gauge theories}

Now let us come back to the study of four-dimensional gauge theories.
It is standard to characterize the phases of gauge theories using the expectation values of line operators.
Let us start by discussing a situation when the theory has a mass gap; i.e.\ there are no massless excitations above the ground state.  If the gauge group is simply connected and all Wilson lines are present, a Higgs phase is characterized by the fact that the expectation values of all the Wilson lines exhibit a perimeter law.  If some Wilson lines have an area law, we say that the theory is confining.

If the gauge group is $G=\tilde G/\bH$ with a nontrivial $\bH$, then some Wilson lines are not present, and therefore we have fewer diagnostics of electric confinement.  But in that case we can use the magnetic and dyonic lines we discussed above to characterize the phases.

As above, we label the lines by their classes $(z_e,z_m)\in \bC\times\bC$, where $\bC$ is the center of $\tilde G$.  Clearly, all the lines in a given class have the same behavior, area law or perimeter law. We refer to a class with an area law as confined.  It is straightforward to multiply these lines.  If $(z_e,z_m)$ and $(z'_e,z'_m)$ are not confined, i.e.\ they have a perimeter law, then so does their product.  If $(z_e,z_m)$ is confined and $(z'_e,z'_m)$ is not, then their product is confined.  But the product of two confined classes might not be confined.  For example, the fundamental Wilson line in an $SU(N)$ pure gauge theory is confined, but its $N$-th power is not.  (More subtle examples, where the $t$-th power of this Wilson line is not confined, where $t$ is a divisor of $N$, were studied in \refs{\CachazoZK,\KonishiTX}.)

It is often the case that the long distance dynamics involves a topological theory.  For example, the theory could be gapped, but there may be an unbroken discrete gauge symmetry at long distance.  One way this can happen is when the original gauge group $G$ is Higgsed, but an unbroken discrete subgroup remains unbroken.  Another possibility, which we will demonstrate shortly, is of an unbroken discrete gauge group of magnetic degrees of freedom. A discrete gauge theory (see \BanksZN\ for a recent review) with gauge group $\Z_k$ has Wilson line operators and surface operators carrying charges $n=0,\cdots,k-1$. As we will see, some of the line operators that have a perimeter law become Wilson lines of this discrete gauge group at long distances.

There are also various Coulomb phases.  They can be ordinary Coulomb, free electric and free magnetic phases (see e.g.\ \IntriligatorAU).  These phases can be easily incorporated into our discussion, but we will not do it here. Additional characteristics of phases were discussed in \CachazoZK, but we will not pursue them here.

\subsec{The phases of $\CN=1$ SYM theories with $\g=su(2)$}

Let us demonstrate this general discussion in the dynamics of $\CN=1$ supersymmetric pure gauge $su(2)$ theories. These theories are characterized by the complex instanton factor
\eqn\etadef{\eta = \Lambda^6 \sim e^{-{8\pi^2 \over g^2} + i \theta}~,}
where $g$ is the Yang-Mills coupling constant (evaluated at a scale which we have set to one), and $\theta$ is the $\theta$-angle. Classically the theory has a $U(1)_R$ symmetry under which the gauginos $\lambda$ are charged, but in the quantum theory this is broken to a discrete subgroup by an anomaly.

The theory with $G=SU(2)$ has two vacua, in which the discrete $\Z_4$ global R-symmetry is spontaneously broken to $\Z_2$. These vacua are characterized by the vacuum expectation value (VEV) of the gaugino bilinear,
\eqn\lambdalambdat{\langle \lambda \lambda \rangle = \pm \eta^{1\over 2}~.}
It is clear from this expression that the two vacua are exchanged under $\theta\to \theta + 2\pi$ (which is equivalent to the $\Z_4$ global symmetry transformation). When this theory is obtained via a mass deformation of the $\CN=2$ pure SYM theory, these two vacua arise from the condensation of a magnetic monopole or a dyon \SeibergRS, and hence the Wilson line in the fundamental representation (and all lines with $(z_e,z_m)=(1,0)$) exhibits confinement in both vacua.

How does this story change in the theory with $G=SO(3)$? First, in this theory we no longer have the $\Z_4$ global symmetry, which is associated with the shift $\theta \to \theta+2\pi$. Instead, the anomaly free  R-symmetry is a $\Z_2$ symmetry, associated with shifting $\theta \to \theta+4\pi$. This symmetry (taking $\lambda \to -\lambda$) is part of the Lorentz group, given by a $2\pi $ rotation in spacetime.  Second, it is clear that the theory still has the same two vacua as the $SU(2)$ theory, but since these two vacua are related by $\theta \to \theta+2\pi$, they are now inequivalent.  The difference between the two vacua can be seen by probing the behavior of the line operators.  The $SO(3)_+$ theory has purely magnetic line operators with charge $(\lambda_e,\lambda_m)=(0,1)$.  In one of the two vacua dyons condense.  Since they have both electric and magnetic charges, these 't Hooft lines have an area law -- they are confined.  In the other vacuum, the condensed particles are purely magnetic.  Hence the same line operators have a perimeter law.  In fact, the charge of the condensed monopole is twice that of the loop operator, and is given by $(\lambda_e,\lambda_m)=(0,2)$.  Therefore, at low energy we find in this vacuum an unbroken $\Z_2$ gauge theory, acting by $\pm1$ on the magnetic line with charge $(\lambda_e,\lambda_m)=(0,1)$ (or, more generally, on all lines with $(z_e,z_m)=(0,1)$). This is an explicit example of our comment above about an unbroken discrete gauge symmetry which appears out of the magnetic degrees of freedom.
The situation is similar in the $SO(3)_-$ theory, except that the two vacua are exchanged.

\subsec{Non-supersymmetric pure Yang-Mills theories with $\g=su(2)$}

We can perform an analogous analysis also for the non-supersymmetric pure gauge theory with gauge groups $SU(2)$ or $SO(3)$. One way to obtain this theory is by adding a gluino mass $m_g$ to the ${\cal N}=1$ supersymmetric theory discussed above, which generically splits the two vacua, and taking the limit
of large $|m_g|$. Consider first the case $|m_g| \ll |\Lambda|$. Since we have a mass gap, the dynamics in each vacuum is essentially the same as above, and their vacuum energy is a positive number times $\pm\Re(m_g\eta^{1\over 2})$  \refs{\EvansHI,\KonishiIZ}. Thus, in the (unique) vacuum of the resulting theory, the $SU(2)$ theory confines (exhibits an area law for its nontrivial line operator), while (depending on the phase of $m_g$ and on the value of the $\theta$-angle) one of the $SO(3)$ theories has a perimeter law for its nontrivial line operator, with an unbroken $\Z_2$ gauge symmetry, while the other $SO(3)$ has an area law. In the $SO(3)$ gauge theory the $\theta$-angle still has periodicity $4\pi$, such that the two theories (and the two low-energy behaviors) are exchanged by $\theta \to \theta+2\pi$.

We conjecture that the same picture holds also in the limit of large $m_g$, where we obtain the pure non-supersymmetric Yang-Mills theory. We do not know how to prove this, since there may be phase transitions as a function of $m_g/\Lambda$. However, let us assume that the standard picture of $SU(2)$ dynamics is correct, namely that the pure $SU(2)$ theory confines (the minimal Wilson line exhibits an area law), and that there is no remaining low-energy discrete gauge symmetry in this case (which seems plausible). Moreover, let us assume that confinement arises from the condensation of some magnetically charged particles (which may or may not carry also some electric charge). Since this condensation is a local phenomenon, it should be the same in the $SU(2)$ and $SO(3)$ theories; in particular both the electric and magnetic charges of the condensing particles must be roots of $su(2)$. The assumption that there is no remaining low-energy discrete symmetry in $SU(2)$ implies that the minimal charge of a condensing particle is not a multiple of any lower charge in the root lattice. However, the form of the charge lattice implies that this charge should either be two times one of the charges of the line operators in the class $(z_e,z_m)=(0,1)$, or two times that of one of the line operators in the class $(z_e,z_m)=(1,1)$. Thus, either the $SO(3)_+$ theory exhibits a perimeter law for its nontrivial line operator and an unbroken $\Z_2$ gauge symmetry, and the $SO(3)_-$ theory exhibits an area law for its nontrivial line operator, or the other way around. Generically we would expect to get one type of behavior for $|\theta| < \pi$, and the opposite behavior for $\pi < |\theta| < 2\pi$. Using the results of the next sections, there are straightforward generalizations of this picture to general $SO(N)$ and $SU(N)/\Z_k$ groups.

The distinction between gapped phases of theories with $\g=su(2)$ is sometimes loosely phrased as a distinction between confining phases, where a monopole condenses, and ``obliquely confining'' phases, where a dyon condenses \tHooftHT. Our discussion above implies a somewhat different distinction. For $G=SU(2)$ one can identify confinement, but one cannot distinguish different confining phases. For $G=SO(3)$ there is no good order parameter for confinement. In this case there are two distinct phases, one with a perimeter law for the ``disorder'' line operator and an unbroken $\Z_2$ discrete gauge symmetry, and one with an area law. However, the relation between
this distinction and the traditional distinction between confining phases is different in $SO(3)_+$ and in $SO(3)_-$, and in any case a local observer cannot tell which of these theories she has (due to \sopm). We suggest that, as in the Higgs/confinement characterization \refs{\BanksFI,\FradkinDV}, the proper distinction between phases should use the behavior of the non-trivial line operators, rather than the identity of the condensed particles.  Furthermore, one should use only the lines that exist in the theory.

\subsec{Compactification on $\S^1$}

Our discussion also has important consequences when the theory is placed on $\R^3 \times \S^1$.  The $\CN=1$ supersymmetric pure $SU(2)$ theory has two vacua in four dimensions, and continues to have two vacua in the compactified theory \SeibergNZ.  However, this is not the case in the $SO(3)_\pm$ theories.  The vacuum with the area law is fully gapped and remains a vacuum in the compactified theory.  But the vacuum with the unbroken $\Z_2$ gauge theory is split to two vacua in the compactified theory.  These two vacua differ by the expectation value of the $\Z_2$ ``Wilson line'' wrapping the $\S^1$ in the low energy theory.  This line is the nontrivial line operator of the microscopic $SO(3)$ theory.  This is an explicit example of the phenomenon mentioned above, that the number of supersymmetric vacua on $\R^3\times \S^1$ (and also the Witten index) depends on the global properties.  More details about this splitting of vacua will appear in \ARSW.

\subsec{Outline}

The rest of the paper is split into two parts.
In sections 2-5, we generalize the discussions so far about $\g=su(2)$ to
 other gauge groups, and analyze the consequences for dualities of supersymmetric gauge theories. We provide ample examples to illustrate the analysis.  Theories with $\g=E_6$, $E_7$ behave almost exactly the same as those with $\g=su(3)$, $su(2)$, respectively, so we do not analyze these cases separately.
In section 6, we will re-analyze our problem from the point of view of the Euclidean path integrals.
These two parts can be read almost separately.

More detailed contents of the paper are as follows.
In section~2, we discuss the lines and the dynamics of gauge theories with $\g=su(N)$. When $G=SU(N)/\Z_N$, we see that the different choices of line operators are permuted by $\theta\to\theta+2\pi$. However, this is not the case for general $SU(N)/\Z_k$ groups. We discuss the properties of the $N$ confined vacua of the $\CN=1$ SYM theory, and the S-duality properties of the $\CN=4$ SYM theory.

In section~3, we similarly discuss the  properties of gauge theories with $G=Spin(N)$ and $G=SO(N)$.
We see that with $N\ge 5$ there is a choice of line operators which cannot be obtained from the naive one by a shift of the standard $\theta$-angle, and we discuss the implications for the pure $\CN=1$ SYM theory, and for the IR duality of SQCD theories.

In section~4, we work out the properties of gauge theories with $\g=sp(N)$.
We find the theories we denote by $G=Sp(N)$, $G=(Sp(N)/\Z_2)_+$ and $G=(Sp(N)/\Z_2)_-$.
We discuss $\CN=4$ theories with these gauge groups and their S-duality properties, and in particular how they map to $\CN=4$ theories with $\g=so(2N+1)$.

In section~5 we study the peculiarities of theories with $\g=so(N)$ with even $N$. In this case,
the choices $N=4d+2$, $N=8d$, and $N=8d+4$ give rise to slightly different behaviors.

In section~6, we show how these choices of line operators can be seen from the point of view of the Euclidean path integral. We see that there are discrete analogues of $\theta$-angles which we can add to the Lagrangian, which reproduce the possible mutually-local sets of line operators. We also indicate how the analysis in the previous sections can be generalized to product gauge algebras. Finally, we study the  surface operators associated to the center and to $\pi_1$ of the gauge group, and discuss how the different theories we discuss (including theories with the same algebra but different gauge groups) may be related by coupling them to discrete $\Z_k$ gauge theories.

\newsec{$su(N)$ gauge theories}

Let us start by considering gauge theories with gauge algebra $\g=su(N)$.
Here the center is $\bC=\Z_N$, so the possible gauge group is $G=SU(N)/\bH$ with $\bH \subset \bC = \Z_N$, namely $\bH=\Z_k$ with $k$ a divisor of $N$.
If the gauge group is $SU(N)$, the allowed purely electric line operators are in $(z_e,z_m)=(n,0)$ with arbitrary $n=0,\cdots,N-1$.  Clearly, the locality condition \lcalr\ shows that all the line operators are in these classes.  In other words, all the magnetically charged lines must be associated with the magnetic root lattice. The situation is more interesting when $\bH$ is nontrivial. We study various choices of $\bH$ in turn.

\subsec{$SU(N)/\Z_N$}
\subseclab\sunzn

Consider first the special case $\bH=\bC=\Z_N$.  Now the purely electric line operators must be in $(z_e,z_m)=(0,0)$, and additional classes of line operators are possible.

\midinsert
\tabskip1em plus1fill\relax
\halign to\hsize{  \hfil #  \hfil & \hfil # \hfil & \hfil # \hfil & \hfil # \hfil\cr
\epsfxsize2.5cm\epsffile{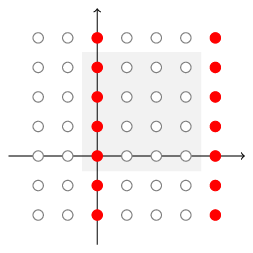} &
\epsfxsize2.5cm\epsffile{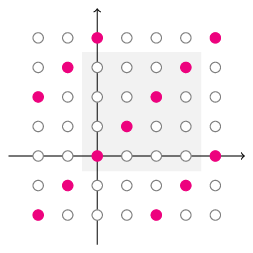} &
\epsfxsize2.5cm\epsffile{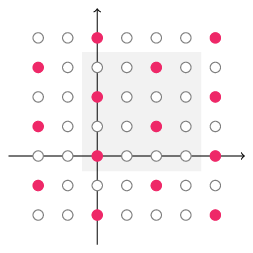} &
\epsfxsize2.5cm\epsffile{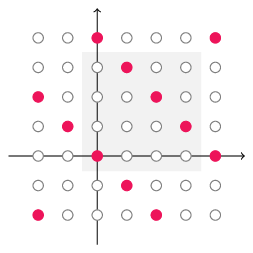} \cr
$(SU(4)/\Z_4)_0$ & $(SU(4)/\Z_4)_1$  & $(SU(4)/\Z_4)_2$ & $ (SU(4)/\Z_4)_3$ \cr}
\figcaption\sufourzfourlines{The $\Z_4$ charges of line operators in the theories $(SU(4)/\Z_4)_{0,1,2,3}$.}
\endinsert

If $N$ is prime, every operator with $z_m\not=0$ can be raised to a power to find an operator with $z_m=1$.  So, without loss of generality, we can assume that a line with $(z_e,z_m)=(n,1)$ exists. Any other line $(z_e,z_m)=(n',1)$ with $n'\not= n$ then does not satisfy \lcalr, and cannot be present. Multiplying the operators in $(z_e,z_m)=(n,1)$ leads to operators in the set
\eqn\sunno{ L_n=\{ (z_e,z_m)=(n m, m)\ \mod\ N\}~}
with $m=0,1,\cdots,N-1$.
Clearly, no other classes of operators can exist.  Hence, for every $n=0,1,\cdots,N-1$ we have a distinct theory $(SU(N)/\Z_N)_n$,  whose line operators have charges in $L_n$ given by  \sunno. This generalizes the case of $SO(3)=SU(2)/\Z_2$ that we discussed above.
The charges of line operators of $(SU(4)/\Z_4)_{0,1,2,3}$ are shown in \sufourzfourlines\ to illustrate the lattices $L_n$.

If $N$ is not prime, the sets $L_n$ of line operators are still valid choices, but are there additional possibilities?  For example, for $N=4$ we can also have a theory with line operators in the classes
\eqn\sunnano{(z_e,z_m)=(0,0), \quad  (0, 2)  ~.}
But we can still add to this list additional line operators without violating mutual locality.
In a consistent quantum field theory, the set of line operators has to be maximal and complete.\foot{Such a completeness requirement is familiar in two-dimensional field theory, and it often follows from imposing modular invariance.
Similarly, our discussion of the Euclidean path integral in section 6 implies that if we put a theory with a non-maximal spectrum of line operators on $\T^4$, it would not be invariant under modular transformations of $\T^4$.}
Then, if the gauge group is really $SU(N)/\Z_N$ so that no lines $(z_e\neq 0, z_m=0)$ are present, there is necessarily a line operator with $z_m=1$. We conclude that even for $N$ that is not prime, the only theories whose gauge group is precisely $G=SU(N)/\Z_N$ are the ones with line operators in one of the sets $L_n$ \sunno.

The $N$ distinct theories that we found have an interesting relation to the $\theta$ parameter.  Because of the Witten effect, the electric charges of magnetically charged particles are shifted as a function of $\theta$.  In our case, the electric weight $\lambda_e$ is shifted by (see \HenningsonHP)
\eqn\lambdaes{\lambda_e \to \lambda_e + \lambda_m ~,}
when $\theta \to \theta + 2\pi$.  In terms of our labels with $(z_e,z_m)$ this means that
\eqn\zes{z_e \to z_e + z_m ~.}
In the $SU(N)/\Z_N$ theories, the sets $L_n$ of operators \sunno\ are transformed as $L_n \to L_{n+1}$.

We see that the shift of $\theta$ by $2\pi$ does not permute the operators in a given $(SU(N)/\Z_N)_n$ theory.  Instead, it permutes the different theories.  We can label the distinct theories by $\theta \in [0,2\pi)$ and $n=0,\cdots,N-1$ as above, or equivalently, we can label them by $\theta \in [0,2\pi N)$
\eqn\SUNL{(SU(N)/\Z_N)_{n}^{{\theta+2\pi}} = (SU(N)/\Z_N)_{{(n+1)\, \mod\, N}}^{\theta}~.}

The fact that the $SU(N)/\Z_N$ theory has an extended range of $\theta$ is known \VafaTF, and it is associated with the existence of $SU(N)/\Z_N$ bundles on spin manifolds with fractional instanton number $\ell \in {1\over N}\Z$.  But it is usually dismissed as a subtlety that is present only when the theory is placed on compact spaces.  Now we see that this extended range of $\theta$ can be detected even when the theory is formulated on $\R^4$ (see also \GaiottoBE).  The spectrum of line operators in the theory detects this subtlety.

\subsec{Dynamics in pure $\CN=1$ SYM theories with gauge groups $SU(N)$ and $SU(N)/\Z_N$}

This understanding of the line operators has significant consequences for the dynamics of the theory.  For concreteness, consider the $\CN=1$ pure SYM theory.  The $SU(N)$ theory has a discrete $\Z_{2N}$ R-symmetry acting on the gauginos, generated by $\gamma$, $\gamma^{2N}=1$, which is associated with a shift of $\theta$ by $2\pi$.  The generator $\gamma^N$ is also in the Lorentz group -- it is a $2\pi$ rotation.  The theory is characterized by the complex instanton factor
\eqn\etadefN{\eta = \Lambda^{3N} \sim e^{-{8\pi^2 \over g^2} + i \theta}~.}
The quantum theory has $N$ vacua, associated with the spontaneous breaking of the global $\Z_{2N}$ symmetry to $\Z_2$.  They are characterized by
\eqn\lambdalambdaN{\langle\lambda \lambda\rangle = \epsilon_N \eta^{1\over N}, ~}
with $\epsilon_N$ an $N$-th root of unity.
The fractional power here signifies that these vacua are permuted by $\theta \to \theta + 2\pi$. The interesting line operators in this theory are the Wilson line in the fundamental representation $W$ in the class $(z_e, z_m)=(1,0)$ and its powers in $(z_e,z_m)=(n ,0)$. 't Hooft line operators exist, but they are associated with roots. $W$ exhibits an area law in all these vacua, signaling confinement.

How does this story change when the gauge group is $SU(N)/\Z_N$?  Clearly, the spectrum of particles, the local operators and their correlation functions in $\R^4$ cannot be modified.  They are invariant under $\theta \to \theta + 2\pi$.  Hence, this theory should still have $N$ vacua that are permuted by this shift.  However, now we know that for a given $\theta \in [0,2\pi)$ there are actually $N$ distinct theories differing by their line operators \sunno.  Since the generator $\gamma$ of the $\Z_{2N}$ symmetry of the $SU(N)$ theory shifts $\theta$ by $2\pi$, and since this operation is not a symmetry of the set of line operators (it maps the theory with $n$ to the theory with $n+1$), the $SU(N)/\Z_N$ theory cannot have a $\Z_{2N}$ R-symmetry. Its only global symmetry is a $\Z_2$ subgroup, which is part of the Lorentz group.

Since the $SU(N)/\Z_N$ theory does not have a discrete symmetry relating its $N$ vacua, the correlation functions of line operators in these vacua can be different. One way to see that this is indeed the case, is to view this $\CN=1$ SYM theory as a mass deformation of the $\CN=2$ pure SYM theory.  The $N$ vacua of the $\CN=1$ theory originate from $N$ multi-monopole points on the moduli space of the $\CN=2$ theory.  The electric and magnetic charges of the monopoles that condense there are the same for $G=SU(N)/\Z_N$ as for $G=SU(N)$. Their electric and magnetic charges are all in the root lattice.

Let us examine the physics of the $N$ vacua.  Without loss of generality we can consider the $SU(N)/\Z_N$ theory with $n=0$.  In this theory  the basic 't Hooft operator $H$ has the charge $(\lambda_e,\lambda_m)=(0,1)$, corresponding to a magnetic weight in the fundamental representation.  Let us first assume that $N$ is prime.  Then, in one of the $N$ vacua the charge of the condensed monopoles is $(\lambda_e,\lambda_m)=(0,N)$, and is aligned with the charges of $H$.  Therefore, $H$ exhibits a perimeter law.  This perimeter law signifies the fact that the magnetic gauge group is Higgsed to $\Z_N$, and at low energies we have a $\Z_N$ gauge symmetry. In the remaining $N-1$ vacua the condensed dyons carry nontrivial electric charges, and therefore $H$ exhibits an area law, as do all the other line operators carrying nontrivial center charges.

If $N$ is not prime, the discussion is a bit more involved.  Consider the $k$-th vacuum.  By shifting $\theta$ by $2\pi k$, the condensed dyons in that vacuum become purely magnetic.  After that shift the nontrivial line operator $H$ belongs to the class $(z_e,z_m)=(k,1)$. For $k\neq 0$ this line still exhibits an area law.  However, the line operator $H^{N/l}$ with $l=\gcd(N,k)$ belongs to the class $(z_e,z_m)=(0,N/l)$, and therefore its charges are aligned with those of the condensed monopoles.  Hence, $H^{N/l}$ exhibits a perimeter law signaling the breaking of the magnetic gauge symmetry to $\Z_l$.  We conclude that in the $k$-th vacuum there is an unbroken $\Z_{l}$ gauge symmetry. The line operator associated with this symmetry is $H^{N/l}$, and we will discuss the surface operators related to it in section 6.4 below.

This understanding of the $N$ vacua leads to interesting consequences when the theory is placed on $\R^3 \times \S^1$.  If the gauge group is $SU(N)$, the four-dimensional theory has $N$ confining vacua, which are related by a symmetry.  The compactified theory has the same $N$ vacua.  However, if the gauge group is $SU(N)/\Z_N$, some of the vacua have an unbroken $\Z_l$ gauge symmetry.  Such a vacuum leads to $l$ vacua in the compactified theory.
So, the total number of supersymmetric vacua of $SU(N)/\Z_N$ SYM theories on a circle is $\sum_{k=1}^N \gcd(N,k)$. For prime values of $N$, this is $2N-1$.

\subsec{$SU(N)/\Z_k$ with $k$ a divisor of $N$}

\topinsert
\tabskip1em plus1fill\relax
\halign to\hsize{  \hfil #  \hfil & \hfil # \hfil & \hfil # \hfil \cr
\epsfxsize2.5cm\epsffile{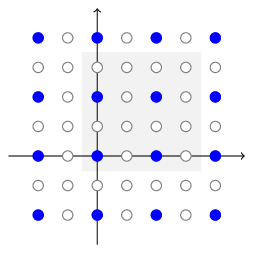} &
\epsfxsize2.5cm\epsffile{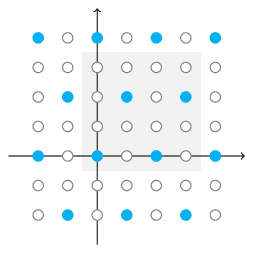} &
\epsfxsize2.5cm\epsffile{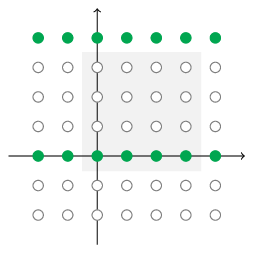} \cr
$(SU(4)/\Z_2)_0$ & $(SU(4)/\Z_2)_1$  & $SU(4)$  \cr}
\figcaption\sufourztwolines{The $\Z_4$ charges of line operators in the theories $(SU(4)/\Z_2)_{0,1}$ and $SU(4)$.}
\endinsert

Let us classify the allowed sets of charges of line operators when the gauge group is  $G=SU(N)/\Z_k$.
Let us write $kk'=N$.  The purely electric line operators have the charges proportional to $(z_e,z_m)=(k,0)$ mod $N$.
The line operator with minimal magnetic charge has the charge $(z_e,z_m)=(n,k')$ mod $N$ for some $n$;  clearly $n$ can be chosen from $0$ to $k-1$. Our completeness requirement implies that one, and exactly one, of these line operators must appear.
The locality does not place any further condition on $n$.
Thus, we see that the allowed choices of the sets of charges of line operators are
\eqn\sunnok{ L_{k,n}=\{ (z_e,z_m)=e(k,0)+m(n ,k')\ \mod\ N\},~}
where $e$ and $m$ are integers, and $n=0,1,\ldots,k-1$.
With this extended notation, the set $L_n$ in \sunno\ is $L_{N,n}$.
We denote by $(SU(N)/\Z_k)_n$ the $SU(N)/\Z_k$ theory with line operators given by $L_{k,n}$.
As examples, we show the cases $(SU(4)/\Z_2)_{0,1}$ in \sufourztwolines\ (compare to \sufourzfourlines).

The shift of the $\theta$-angle by $2\pi$ sends $(z_e,z_m)$ to $(z_e+z_m,z_m)$.
Therefore this sends $L_{k,n} \to L_{k,n+k'}$. In other words, we have
\eqn\SUNK{(SU(N)/\Z_k)_{n}^{{\theta+2\pi}} = (SU(N)/\Z_k)_{{n+k'\, \mod\, k}}^{\theta}~.}
When $\gcd(k,k')=1$, this implies that every choice of $n$ can be reached by shifting the standard $\theta$-angle $\theta$. However, when $\gcd(k,k')=l \neq 1$, however, the shift of $\theta$ only maps the theory $(SU(N)/\Z_k)_n$ to other theories with the same ($n$ mod $l$), and there are $l$ sets of theories that are not related by shifts of the $\theta$-angle. This is our first example of theories with the same gauge group, which are distinguished by a ``discrete $\theta$-angle'' that is not related to the original $\theta$-angle; we will discuss the Euclidean path integral realization of this distinction in section 6 below. This can happen if and only if $N$ has some prime factor that appears more than once in its decomposition into primes (namely, $N$ is not square-free). The first example is $N=4$, where the $(SU(4)/\Z_2)_0$ and $(SU(4)/\Z_2)_1$ theories are not related by shifting the $\theta$-angle.

It is possible to generalize our previous analysis of the behavior of the different vacua of the $\CN=1$ pure SYM theory to all these theories, but we will not do this here.

\subsec{S-duality of $\CN=4$ supersymmetric theories}

Let us consider now the transformation under S-duality of $\CN=4$ SYM theories. When the gauge group is $U(N)$, the theory is invariant under an $SL(2,\Z)$ transformation, which maps the complex gauge coupling
\eqn\deftau{\tau = {\theta \over {2\pi}} + {{4\pi i}\over g^2}}
to $\tau \to (a \tau + b) / (c\tau + d)$, with $a,b,c,d$ integers satisfying $ad-bc=1$. This transformation group is generated by $T$ which takes $\theta \to \theta + 2\pi$ or $\tau \to \tau+1$, and by a generator $S$ which takes $\tau \to -1/\tau$.

Let us now consider $\CN=4$ SYM theories with gauge algebra $\g=su(N)$. The magnetic dual algebra in this case is also $\g^*=su(N)$. Usually it is stated that the theory with $G=SU(N)$ is mapped by the $S$ generator to the theory with $G=SU(N)/\Z_N$, but we found that there are $N$ variations of the latter theory, so the full story is more complicated\foot{The case of $\g=su(2)$ was discussed in \refs{\GukovJK,\GaiottoBE}.}.
In general we found above that there are distinct theories $(SU(N)/\Z_k)_{n\ \mod\ k}$, differing by the global structure of the gauge group and by the set of charges of the line operators $L_{k,n}$.

Let us consider how the S-duality acts on this set of theories\foot{In theories with extended supersymmetry there are also BPS versions of line operators and surface operators, which are labeled by additional parameters. We do not discuss these here.}. We already discussed above how the $T$ generator maps a given theory to another theory with the same $k$, but with a possibly different value of $n$.
The $S$ generator acts on the weights $(\lambda_e,\lambda_m)$ of the line operator as
\eqn\dualweights{S:\quad (\lambda_e,\lambda_m) \quad\mapsto\quad (\lambda_m,-\lambda_e).}
Therefore, it takes the set $L$ of conjugacy classes of charges of line operators to
the set $L^*$ of charges of the dual theory, given by
\eqn\dualcharges{(z_e,z_m)\in L \quad \Longleftrightarrow\quad  (z_m,-z_e)\in L^*.}
For example, when the original theory is $(SU(N)/\Z_k)_0$ with the set $L_{k,0}$ of charges,
the dual set of charges is precisely $L_{k',0}$ where $kk'=N$. Therefore the dual theory is $(SU(N)/\Z_{k'})_0$. This is what is usually stated as the change in the global structure of the gauge group under S-duality.

It is not difficult to generalize this to the full set of theories that we discussed above.
Given $L_{k,n}$, the lattice $(L_{k,n})^*$ must again be of the form $L_{k^*,n^*}$, determining the S-dual of $(SU(N)/\Z_k)_n$ to be $(SU(N)/\Z_{k^*})_{n^*}$.
The numbers $k^*$ and $n^*$ can be determined by finding the minimal charges of the form
\eqn\dualminimal{	(0,k^*),  \qquad (-N/k^*,n^*) \in L_{k,n}. }
First, the charge of the form $(0,k^*)$ can be found by choosing $e$ and  $m$ in \sunnok\ to be $e=-n/\gcd(n,k)$, $m=k/\gcd(n,k)$. Here we use the convention $\gcd(0,k)=k$. Then, we have
\eqn\kstar{k^*=kk'/\gcd(n,k)=N/\gcd(n,k).}
We can find integers $e_0$ and $m_0$ such that $e_0k+m_0n=\gcd(k,n)$. Then, $n_*=-m_0k'$.

It is interesting to ask whether the full $SL(2,\Z)$ group maps all these different theories to each other or not. For prime values of $N$ the answer is clearly yes; the $S$ generator maps $SU(N)$ to $(SU(N)/\Z_N)_0$, and the $T$ generator then relates this to all the other theories. However, in general the answer is no. If there is some $k$ such that $\gcd(k,N/k)=l$, then the $(SU(N)/\Z_k)_0$ has electric and magnetic charges that are all multiples of $l$, and this is still the case also after we perform any $SL(2,\Z)$ transformation. Thus, these theories sit in separate orbits of $SL(2,\Z)$ from the $SU(N)$ and $SU(N)/\Z_N$ theories.

In fact, this is the only case where separate orbits exist; namely, if $N$ is square-free (every prime factor in the decomposition of $N$ into primes appears once), then all the different theories we discussed are in the same orbit of $SL(2,\Z)$. To see this, note first that the $SU(N)$ gauge theory is mapped to itself under the subgroup of $SL(2,\Z)$ that is generated by $T$ and by $S T^N S$. This subgroup is denoted by $\Gamma_0(N)$. The space of couplings of the $SU(N)$ theory that are inequivalent is thus a fundamental domain of $\Gamma_0(N)$. However, we know that every $SL(2,\Z)$ transformation relates our $SU(N)$ theory to one of the other theories, and that every fundamental domain of $SL(2,\Z)$ contains precisely one value of the coupling which is related by $SL(2,\Z)$ to weak coupling ($\tau \to i\infty$). So, the number of different weak coupling limits that an $SU(N)$ gauge theory has, or, in other words, the number of inequivalent theories that it maps to under the full $SL(2,\Z)$ group, is given by the number of fundamental domains of $SL(2,\Z)$ inside the fundamental domain of $\Gamma_0(N)$. This number, called the index of $\Gamma_0(N)$, is equal to
\eqn\indexgamma{{\rm index}(\Gamma_0(N)) = N \prod_{p | N} (1 + {1\over p}),}
where the product runs over all the prime factors of $N$. On the other hand, the total number of distinct theories that we found is $\sum_{k | N} k$, where the sum goes over all the divisors $k$ of $N$. It is easy to check that these two numbers are the same if and only if $N$ is square-free.
In this case all the different theories we discussed are related by $SL(2,\Z)$ transformations, while otherwise this is not the case. The first value of $N$ exhibiting separate $SL(2,\Z)$ orbits is $N=4$, where the $(SU(4)/\Z_2)_0$ theory is mapped to itself under the full $SL(2,\Z)$ group, while the other $6$ theories are permuted.

\topinsert
\tabskip1em plus1fill\relax
{
\halign to\hsize{  \hfil #   &  \hfil # \hfil \cr
$su(3)$: & $\vcenter{\hbox{\epsffile{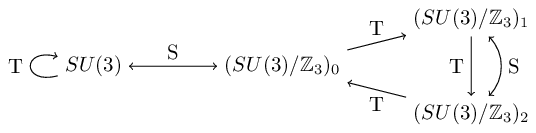}}}$ \cr
$su(4)$: & $\vcenter{\hbox{\epsffile{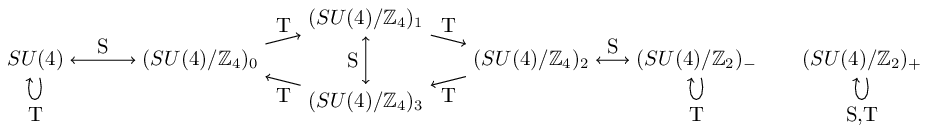}}}$ \cr
$su(6)$: & $\vcenter{\hbox{\epsffile{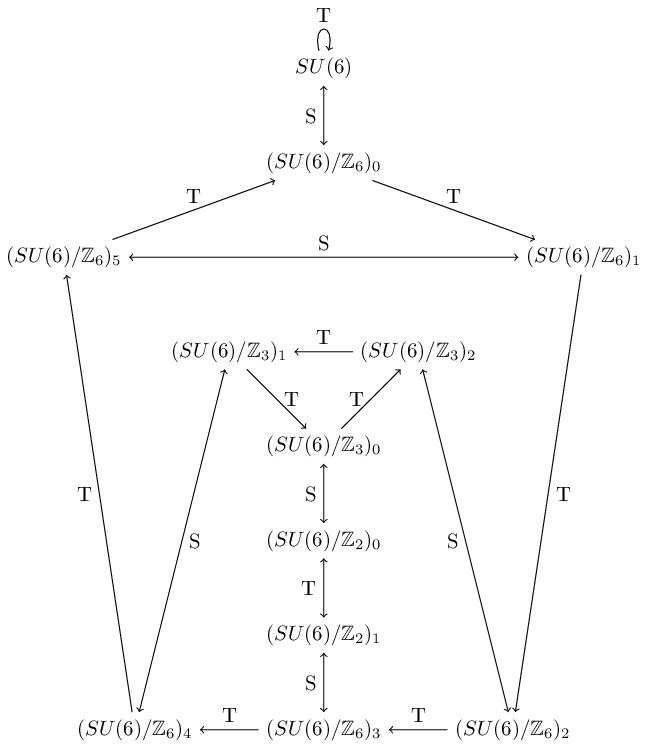}}}$ \cr
}}
\medskip
\figcaption\suorbits{S-duality orbits of the $\CN=4$ SYM theory with $\g=su(3), su(4), su(6)$.}
\endinsert

To illustrate the discussions so far, we show the duality orbits of $\CN=4$ theories with $\g=su(3)$,
$\g=su(4)$ and $\g=su(6)$ in \suorbits.\foot{%
Some of the arrows in the $su(6)$ case were wrongly oriented in previous versions. 
The authors thank Francesco Benini for pointing out the issue.
} 
As discussed, we see that there is just one orbit for $\g=su(3)$ and $\g=su(6)$, but there are two orbits for $\g=su(4)$.

The distinctions between the different theories that we discuss here are
important when computing the partition functions of ${\cal N}=4$ SYM on
compact manifolds, and using them to test S-duality \refs{\VafaTF,\GirardelloGF,\WuBV}. We will discuss the precise way to compute these partition functions in section 6.

The AdS/CFT correspondence \MaldacenaRE\ maps the $\CN=4$ SYM theory with
$\g=su(N)$ to the type IIB string theory on $AdS_5\times S^5$. In this context the
distinction between the different theories we discuss arises from the need
to carefully quantize the topological theory
\eqn\topiib{S = {N\over {2\pi i}}  \int_{AdS_5} B_{RR} \wedge dB_{NS}}
that arises in the type IIB string theory at low energies \refs{\AharonyQU,\WittenWY,\AharonyTI}.
One possibility is to make the $U(1)$ gauge field (dual to
the singleton modes in the bulk, including the two-forms $B_{RR}$ and $B_{NS}$) dynamical, so that the gauge theory becomes a
$U(N)$ theory (see e.g.\ \refs{\MaldacenaSS,\BelovHT}). Similar issues arise also in the $6d$ $A_N$ $(2,0)$ SCFT, dual to M theory on
$AdS_7\times S^4$, where the surface operator \GanorNF\ is not mutually local with respect
to itself. Such theories, containing operators that are not mutually local,
are analogous to holomorphic blocks of $2d$ conformal field theories. They are
not standard consistent local quantum field theories by themselves, but can be viewed either as
generalizations of local quantum field theories where the expectation values are
elements of a vector space rather than well-defined numbers \WittenWY\ (see also \refs{\Segal,\MooreJV,\WittenAT,\FreedBS}, and \mooretalks\ and references therein), or as
building blocks for constructing consistent local quantum field theories.  Consistent string constructions must lead to completely local theories, and this places further constraints on them \SeibergDR.

\newsec{$so(N)$ gauge theories}

\subsec{The theories}
\subseclab\sonsection

For odd $N$ the center $\bC$ of $Spin(N)$ is $\Z_2$, where the $\Z_2$ charge is carried by all the spinor representations. The dual group in this case is $\g^*=sp((N-1)/2)=usp(N-1)$, which also has a $\Z_2$ center carried by the fundamental (vector) representation. The possible lines are in the sets $(z_e,z_m)$ with $z_e=0,1$ and $z_m=0,1$.  The $Spin(N)$ theory has line operators with $(z_e,z_m)=(0,0)$ and $(z_e,z_m)=(1,0)$. There are two different theories with $SO(N)$ gauge group, generalizing our previous discussion for $N=3$. The $SO(N)_+$ theory has line operators with $(z_e,z_m)=(0,0)$ and $(z_e,z_m)=(0,1)$, while the $SO(N)_-$ theory has lines with $(z_e,z_m)=(0,0)$ and $(z_e,z_m)=(1, 1)$.

For even $N$ the dual magnetic group is $\g^*=so(N)$, and the center $\bC$ of $Spin(N)$ depends on $(N\, \mod\, 4)$.  It is $\Z_2\times \Z_2$ for $(N=0\, \mod\, 4)$ and it is $\Z_4$ for $(N=2\, \mod\, 4)$.
Correspondingly, the representations are split into four classes, which are the vector class, the adjoint (trivial) class, and two spinor classes.
In section~5 we will discuss the $Spin(N)$ and $Spin(N)/\bH$ theories for all these cases in detail. Here we will focus on a particular quotient $SO(N)=Spin(N)/\Z_2$, which has electric Wilson lines in the trivial class and in the vector class. Mutual locality requires the lines that are not purely electric to have magnetic charges in the vector class. One can then either choose these lines to be purely magnetic, or dyonic with spinor electric charges. Correspondingly, there are two theories $SO(N)_\pm$, just as in the case when $N$ is odd.

Up to now the situation is very similar to the discussion for $N=3$ around \sopm.  However, the $N=3$ and $N=4$ cases are somewhat different than higher values of $N$.  For $N=3$ we saw that the shift of $\theta$ by $2\pi$ exchanges the two theories.  For $N=4$, we have $SO(4)= [SU(2) \times SU(2)]/\Z_2$, and therefore there are actually two $\theta$-angles, $\theta_s$ with $s=1,2$, one for each $SU(2)$ factor. In this case the Witten effect gives an action on line operators that implies
\eqn\softheta{SO(4)^{{\theta_1,\theta_2}}_{+} = SO(4)^{{\theta_1+ 2\pi,\theta_2}}_{-}~.}

For higher values of $N$ the two $SO(N)$ theories are not related by a shift of $\theta$, but rather
\eqn\sotheta{SO(N)^{\theta}_{\pm} = SO(N)^{{\theta+2\pi}}_{\pm} ~.}
This can be seen by examining the roots and weights of $so(N)$ and its Langlands duals $usp(N-1)$ (for odd $N$) or $so(N)$ (for even $N$), and by examining the shift of the electric charges under a shift of $\theta$.  Equivalently, the same conclusion follows from the fact that these theories do not have half-instantons \WittenNV, so they must be invariant under $\theta \to \theta+2\pi$. Note that in the $SO(4)$ theory a configuration which has a half-integer instanton number in the first $SU(2)$, and a half-integer instanton number in the second $SU(2)$, can be present as long as the sum of the two instanton numbers is an integer.

\subsec{The pure $\CN=1$ SYM theory}

As in the previous examples, we consider now the implications for the $\CN=1$ pure SYM theory.  The $Spin(N)$ theory with $N>4$ has a discrete $\Z_{2(N-2)}$ R-symmetry acting on the gauginos. The theory is characterized by the instanton factor $\eta = \Lambda^{3(N-2)}$.
It has $(N-2)$ vacua with
\eqn\lambdalambdaso{\langle \lambda\lambda \rangle = {1\over 2} (16 \eta)^{1\over N-2} \epsilon_{N-2} ~,}
with $\epsilon_{N-2}$ an $N-2$-th root of unity.  (We use here the conventions of \refs{\IntriligatorID,\IntriligatorAU}.)  These vacua are associated with the breaking of $\Z_{2(N-2)}$ to $\Z_2$, which is in the Lorentz group.

Again, $N=3$ and $N=4$ are slightly different.  We have already discussed $N=3$ above.  For $N=4$, $Spin(4)=SU(2)\times SU(2)$ so the global symmetry is $\Z_4 \times \Z_4$ (a separate $\Z_4$ for each factor), and there are four vacua with $\langle(\lambda\lambda)_s\rangle = \epsilon_2(s) \eta_s^{1\over 2} $, where $s=1,2$ labels the two $SU(2)$ factors and $\eta_s=\Lambda_s^6$ are their instanton factors.  The effective superpotential in these vacua is
\eqn\effsup{W=2 \eta_1^{1\over 2}\epsilon_2(1) +2\eta_2^{1\over 2}\epsilon_2(2) ~.}
Something special happens for $\eta_1= \eta_2$.  Then, in the two vacua with $\epsilon_2(1)=\epsilon_2(2)=\pm 1$ the superpotential is nonzero, and in the other two with $\epsilon_2(1)=-\epsilon_2(2)=\pm 1$ the superpotential vanishes.  This point was crucial in \refs{\IntriligatorID,\IntriligatorAU}\ and will be important below.

Both for even and odd $N$, the nontrivial Wilson line in the spinor representation exhibits an area law in all of these vacua.  For even $N$ the Wilson line in the vector representation is also confined. (For odd $N$ it is screened by the gluons, because they are in the same conjugacy class.)

For $N > 4$, the $SO(N)_\pm$ theories are invariant under $\theta \to \theta +2\pi$, so they also have the $\Z_{2(N-2)}$ symmetry and $N-2$ vacua.  But the dynamics of the vacua are different.  Using the notation of odd $N$ (with an obvious interpretation for even values of $N$)
in all the vacua of the $SO(N)_+$ theory the nontrivial line with $(z_e,z_m)=(0,1)$ has a perimeter law associated with an unbroken $\Z_2$ gauge symmetry.  On the other hand, in all the vacua of the $SO(N)_-$ theory the nontrivial line with $(z_e,z_m)=(1,1)$ has an area law. For even values of $N$, the Wilson line in the vector representation also has an area law in all these cases.

Correspondingly, the $Spin(N)$ theory on $\R^3\times \S^1$ has $(N-2)$ vacua.  The $SO(N)_+$ theory on $\R^3\times \S^1$ has $2(N-2)$ vacua, while the $SO(N)_-$ theory has $(N-2)$ vacua.

Again, the case of $N=4$ is slightly different.  In the $SO(4)_\pm$ theories, the global symmetry\foot{Note that the symmetry group in this case is not a standard supersymmetry algebra with an R-symmetry and a global symmetry that commutes with supersymmetry; this is related to the fact that the local dynamics of these theories is a sum of separate $SU(2)$ theories, with separate super-Poincar\'e symmetries acting on each one.} is not $\Z_4\times \Z_4$ symmetry but rather $\Z_4 \times \Z_2$.  The four vacua of the $Spin(4)$ theory, labeled by $\epsilon_2(s)=\pm 1$, are still present.  In the $SO(4)_+$ theory, the two vacua with $\epsilon_2(1)=\epsilon_2(2)=\pm 1$ exhibit a perimeter law for the nontrivial 't Hooft line operator, associated with an unbroken $\Z_2$ gauge symmetry. In the other two vacua, that have $\epsilon_2(1)=-\epsilon_2(2)=\pm 1$, this line operator is confined. This situation is reversed in the $SO(4)_-$ theory. The Wilson line in the vector representation is confined in all of these cases. Correspondingly, the two $SO(4)$ theories both have $6$ supersymmetric vacua on $\R^3\times \S^1$.

\subsec{$\CN=1$ with vectors -- duality}

We saw above that when we include in the spectrum of line operators the Wilson line in the vector representation, the discussion of odd and even values of $N$ is very similar. This is the case in particular whenever we have dynamical fields in the vector representation, which can screen the
Wilson lines in the vector representation.

The dynamics of the $\CN=1$ supersymmetric $SO(N)$ gauge theory for different numbers of colors $N$ and flavors $N_f$ was analyzed in detail in \refs{\IntriligatorID,\IntriligatorAU,\IntriligatorER}. In that discussion the focus was on the local structure. We saw above that each of these theories comes in three versions: $Spin(N)$, $SO(N)_+$, and $SO(N)_-$. Here we will extend the analysis to take this fact into account, focusing on the case of $N_f \geq N$.

 For these values of $N_f$ it was found that the $so(N)$ theory with $N_f$ chiral multiplets $Q_i$ ($i=1,\cdots,N_f$) in the vector representation and no superpotential is dual at low energies to an $so(N_f-N+4)$ theory with $N_f$ chiral multiplets $q_i$ in the vector representation, singlet mesons $M^{ij}$ ($i,j=1,\cdots,N_f$), and a  tree level superpotential \refs{\SeibergPQ,\IntriligatorID,\IntriligatorAU}
\eqn\dualsup{W={1\over 2\mu} M^{ij} q_i q_j~.}
The parameter $\mu$ is related to the instanton factors $\eta$ and $\tilde \eta$ of the two theories through
\eqn\etaetat{\eta \tilde \eta= {1\over 2^8}(-1)^{N_f-N_c} \mu^{N_f}. }
We claim that taking the global structure into account this duality actually maps
\eqn\dualitymap{\eqalign{
&Spin(N) \longleftrightarrow SO(N_f-N+4)_- \cr
&SO(N)_+ \longleftrightarrow SO(N_f-N+4)_+ \cr
&SO(N)_- \longleftrightarrow Spin(N_f-N+4) \cr}
}
 The discussion in  \StrasslerFE\ can be interpreted as indicating a duality between $Spin(N)$ and $SO(N_f-N+4)$, but it did not take into account the two different $SO(N)$ theories, and therefore did not discuss the $SO(N)_+ \longleftrightarrow SO(N_f-N+4)_+$ duality. Note that, for $N, N_f-N+4 > 4$, the periodicity of the $\theta$-angle in all three theories is $2\pi$ \sotheta, consistent with \etaetat; we will discuss what happens for lower values of $N$ and $N_f-N+4$ in the next subsection.

We start with the mapping for $N_f \geq N > 4$; one can flow from here to lower values of $N_f$ and $N$ by turning on mass terms and Higgsing. The original $so(N)$ theory has a moduli space of vacua labeled (partly) by $M^{ij}=Q^i Q^j$, which satisfies ${\rm rank}(M) \leq N$. Consider the component of the moduli space where ${\rm rank}(M)=N$, so that the gauge group is completely broken. For every such value of $M$ we have two supersymmetric vacua, differing by the sign of the vacuum expectation value of a baryon operator $B=Q^N$. For large $\vev{M}$ the theory is weakly coupled, and it is clear that in the $Spin(N)$ theory, the spinor Wilson line has a perimeter law (related to a $\Z_2$ gauge symmetry), while the nontrivial line operators of the $SO(N)_\pm$ theories exhibit an area law.

In the dual theory, with the same expectation values of $M$, $N$ of the flavors become massive, so at low energies there is an $so(N_f-N+4)$ theory with $(N_f-N)$ massless flavors $q$. Ignoring the superpotential for a moment, we can give expectation values to these flavors, and break the gauge group to $so(4)$, with equal instanton factors $\eta_{low}$ for its two $su(2)$ factors, proportional to powers of the various VEVs.
Now we can repeat the analysis around \effsup\ of this low energy effective theory.  The theory has four vacua.  Two of them, with $\epsilon_2(1)=\epsilon_2(2)=\pm 1$ have a nonzero superpotential proportional to $\sqrt{\eta_{low}}$, which in our case leads to a runaway behavior with no supersymmetric vacuum. The other two vacua, the ones with $\epsilon_2(1)=-\epsilon_2(2)=\pm 1$, have a vanishing superpotential, and we can identify them with the two vacua that we found above \IntriligatorID\ (for a given expectation value of $M$, and after fixing the expectation value of $q$ using the superpotential).  When the gauge group is $Spin(N_f-N+4)$, the nontrivial line operator is a spinor Wilson line, which is confined in these vacua.
When the gauge group is $SO(N_f-N+4)_\pm$ there is also a single nontrivial line operator. Our discussion of the line operators of $SO(4)$ at the end of section 3.2 implies that this line operator has an area law in the $SO(N_f-N+4)_+$ theory, and a perimeter law in the $SO(N_f-N+4)_-$ theory, where there is an unbroken gauged $\Z_2$ symmetry as discussed above. We conclude that the $Spin(N)$ theory must map to the $SO(N_f-N+4)_-$ theory. Since performing the duality twice should lead us back to the original theory, we find the mapping \dualitymap.

We can test this result by turning on masses for all the flavors of the $so(N)$ theory. In the original theory we flow at low energies to the $(N-2)$ vacua of the pure $so(N)$ SYM theory, discussed in the previous subsection. As we discussed there, the $Spin(N)$ and $SO(N)_-$ theories exhibit an area law for their nontrivial line operator, while the $SO(N)_+$ theory exhibits a perimeter law.

Let us check what happens in the dual theory \refs{\IntriligatorID,\IntriligatorAU}, which now has the tree level superpotential
\eqn\dualsupm{W={1\over 2\mu} M^{ij} q_i q_j + {1\over 2} m_{ij}M^{ij}~,}
 where the second term in \dualsupm\ is the mass term we turned on in the original $so(N)$ theory. We should find the $(N-2)$ vacua of the original theory in this theory.

 Classically, the equations of motion of $M^{ij}$ from \dualsupm\ lead to $q_i q_j =- \mu m_{ij}$, so we must have nonzero VEVs for $q_i$ such that $q_i q_j$ has rank $N_f$, but this is impossible since $N_f > N_f-N+4$. Thus, there are no classical vacua.
 To find supersymmetric vacua, let us explore the region in field space with generic nonzero $M$. Here the dual quarks $q_i$ are massive and they can be integrated out.  The low energy theory is a pure gauge $so(N_f-N+4)$ theory with scale $\tilde \eta_{low} = \tilde \eta \mu^{-N_f} \det(M)$.  Gaugino condensation in this pure gauge theory leads to an effective superpotential
\eqn\Weffma{W_{eff}= {1\over 2 } (N_f-N+2)(16 \mu^{-N_f} \tilde \eta \det (M))^{1/(N_f-N+2)} + {1\over 2} m_{ij}M^{ij}~.}
The equations of motion of $M$ now lead to $(N-2)$ supersymmetric vacua with
\eqn\Mvevm{
\eqalign{
\langle M^{ij}\rangle & = \epsilon_{N-2} \left({(-1)^{N_f-N} \mu^{N_f}\det (m) \over {16 \tilde \eta}}\right)^{1\over {(N-2)}} \left({1\over m}\right)^{ij} \cr
& = \epsilon_{N-2} \left(16 \eta \det (m) \right)^{1\over {(N-2)} } \left({1\over m}\right)^{ij}~,}
}
where $\epsilon_{N-2}$ is an $(N-2)$-th root of one. These $(N-2)$ vacua map exactly (including the value of the superpotential) to those of the original theory. In these vacua again the nontrivial line operators of the $Spin(N_f-N+4)$ and $SO(N_f-N+4)_-$ theories exhibit an area law (as in our analysis of the previous subsection), while that of the $SO(N_f-N+4)_+$ theory exhibits a perimeter law. This is consistent with our mapping \dualitymap.

The mapping \dualitymap\ is also consistent with the relation between the ${\cal N}=1$ duality and the low-energy dynamics of ${\cal N}=2$ supersymmetric gauge theories \refs{\ArgyresEH,\ArgyresHC} (see also \refs{\ShifmanYH,\ShifmanEWA} and references therein). The moduli space of the ${\cal N}=2$ SQCD theory with $\g=so(N)$ and $N_f/2$ fundamental hypermultiplets contains (for $N_f < 2N-4$) a point where the low-energy theory is an $\g=so(N_f-N+4)$ theory with $N_f/2$ fundamental hypermultiplets. Upon breaking to ${\cal N}=1$ supersymmetry by adding a mass term for the adjoint chiral superfield, the physics at this point has two different descriptions, which realize a deformation of the ${\cal N}=1$ duality described above by a quartic superpotential for the quarks. In this realization, the $so(N_f-N+4)$ theory originates
from a subgroup of the $so(N)$ theory, so the allowed electric weights of line operators in the $so(N_f-N+4)$ theory should be a subset of the allowed weights in the $so(N)$ theory (when viewing $so(N_f-N+4)$ as a subalgebra of $so(N)$). This is consistent with our mapping \dualitymap, that maps all the lines with spinorial electric charges to other lines with spinorial electric charges.

\subsec{Triality for $N_f=N-1$ and for $N_f=N$}

In the cases of $N_f=N$ and $N_f=N-1$, we obtain $so(3)$ and $so(4)$ theories on the right-hand side of \dualitymap, and in these theories the $SO(N)_+$ and $SO(N)_-$ theories are related by shifting the $\theta$-angle, as we saw in \sopm, \softheta. This gives us an extra relation also between the $so(N)$ theories on the left-hand side; in \refs{\IntriligatorID,\IntriligatorER} this was called a triality for the case of $N_f=N-1$. In our language it generally involves relations between $4$ different theories.

Let us start with the case of $so(N)$ with $N_f=N-1$. According to \IntriligatorID, the dual description
of this is as above, but with an extra term in the superpotential;
this theory is equivalent at low energies to an $so(3)$ theory with singlets $M$ and with
\eqn\sothreepot{W = {1\over {2\mu}} M^{ij} q_i q_i - {1\over {64 \eta}} \det(M),}
and with a scale
\eqn\scalesothree{\tilde\eta = {\mu^{2(N-1)} \over {2^{14} \eta^2}}.}
For example, the $Spin(N)$ theory with $N_f=N-1$ is equivalent to the $SO(3)_-$ theory with this superpotential. But, we know that the latter theory is also equivalent to an $SO(3)_+$ theory with $\tilde\theta$ shifted by $2\pi$. By the duality \dualitymap, and using the relation \scalesothree, this is equivalent to an $SO(N)_+$ theory in which we shift the $\theta$-angle by $\pi$,
taking $\eta\to -\eta$, except that this shift would give us an extra minus sign in the last term of \sothreepot, which we do not get here. So, we find \IntriligatorID\ that the $Spin(N)$ theory with $N_f=N-1$ and with coupling constant $\eta$ is equivalent at low energies to an $SO(N)_+$ theory with coupling $(-\eta)$ and with an extra superpotential
\eqn\trialw{W = -{1\over {32 \eta}} \det(M).}
The spinor Wilson line of the $Spin(N)$ theory is mapped by this duality to the 't Hooft line of the $SO(N)_+$ theory. Similarly, since the shift by $2\pi$ takes the $Spin(3)$ theory to itself, we find that the $SO(N)_-$ theory with coupling $\eta$ is equivalent to another $SO(N)_-$ theory with coupling $(-\eta)$, and with the same superpotential \trialw. Despite appearances, and even though the nontrivial dyonic line of the $SO(N)_-$ theory maps to itself, this is a strong/weak coupling duality, as can be seen by mapping the supersymmetric vacua \IntriligatorID. In particular, vacua in which electrically-charged particles condense are exchanged with vacua in which magnetically-charged particles condense.

We can now repeat the same story for $N_f=N$, where the dual theory has only the first term in the superpotential \sothreepot. The complication here is that to relate the two different $so(4)$ theories we need to shift the $\theta$-angle of one of the $su(2)$ subgroups of $so(4)$ by $2\pi$ (see \softheta), and this does not map just to a change of the $\theta$-angle in the original theory. Using the fact that the difference between the operators $(W_{\alpha}^2)$ of the two $su(2)$ groups maps to the baryon operator $B = Q^{N}$ (with the color and flavor indices all contracted anti-symmetrically) \IntriligatorID, similar arguments to those of the previous paragraph imply that the $Spin(N)$ theory with $N_f=N$ flavors and coupling $\eta$ is equivalent at low energies to an $SO(N)_+$ theory with coupling $(-\eta)$ and with an extra superpotential
\eqn\trialwn{W = \beta B.}
A similar relation holds for two $SO(N)_-$ theories. It is not clear how to directly find the value of $\beta$, since the operator mapping is only known near the point $\beta=0$. However, we can find this coefficient by adding a mass term, and flowing to the duality described in the previous paragraph; upon integrating out the massive quark, the superpotential proportional to $B$ becomes precisely the superpotential proportional to $\det(M)$ of \trialw. This consideration implies that $\beta = \pm i / 4 \sqrt{\eta}$; the two choices are related by charge conjugation (which, for $so(4)$, exchanges the two $su(2)$ factors).

\newsec{$\g = sp(N)$}

\subsec{The theories}

In this section we consider theories based on the Lie algebra $\g=sp(N) \equiv usp(2N)$.  Its center is $\bC=\Z_2$, with the fundamental (vector) representation charged under the center, so we will be interested in two gauge groups $Sp(N) \equiv USp(2N)$ and $Sp(N)/\Z_2$. The dual algebra is $\g^*=so(2N+1)$, with a nontrivial center representation carried by its spinor weights.

The classes of line operators are labeled by two elements of $\Z_2$: $(z_e,z_m)$.
If the gauge group is $Sp(N)$, the allowed lines are in $(z_e,z_m)=(0, 0)$ and $(z_e,z_m)=(1,0)$.
When the gauge group is $Sp(N)/\Z_2$, the lines in $(z_e,z_m)=(1,0)$ are not present.  Instead, we have two options: add the lines in $(z_e,z_m)=(0,1)$ or add the lines in $(z_e,z_m)=(1,1)$.  We refer to the first option with the purely magnetic lines as $(Sp(N)/\Z_2)_+$, and to the second option with the dyonic line as $(Sp(N)/\Z_2)_-$.  These options with this terminology are consistent with our discussion of $Sp(1)/\Z_2=SO(3)$ around \sopm, and with our discussion of $Sp(2)/\Z_2=SO(5)$ in section \sonsection.

Let us study how the shift of $\theta$ by $2\pi$ affects these options.
The weight lattice of $sp(N)$ is $\Z^N$. An element $\lambda_e=(v_i)\in \Z^N$ is in the adjoint class if the sum of $v_i$ is even. Otherwise it is in the vector class. In the dual $so(2N+1)$ we use a normalization of the weights which is twice the natural normalization; this is needed in order for the shift of $\theta$ to act in a nice way, given that there is a factor of $2$ between the length of the short and long roots \refs{\GirardelloGF,\ArgyresQR}. The magnetic weight lattice is then a sub-lattice of $\Z^N$. An element $\lambda_m=(w_i)\in \Z^N$ is in the magnetic weight lattice if all the $w_i$ are even or if all of them are odd. In the first case it is in the adjoint class, and in the second case it is in the spinor class.
The line with nonzero magnetic charge of the $Sp(N)/\Z_2$ theory has the magnetic weight $\lambda_m=(1,1,\cdots,1)$. In our current normalization the shift of $\theta$ by $2 \pi$ changes $\lambda_e\to \lambda_e+\lambda_m$.
Since $\lambda_e=(1,1,\cdots,1)$ is in the trivial class for even $N$, and in the vector class for odd $N$, we see that
\eqn\spodd{\eqalign{
&(Sp(N)/\Z_2)^{\theta}_{+} = (Sp(N)/\Z_2)^{{\theta + 2\pi}}_{-} \qquad {\rm for\ odd} \ N, \cr
&(Sp(N)/\Z_2)^{\theta}_{\pm} = (Sp(N)/\Z_2)^{{\theta + 2\pi}}_{\pm} \qquad {\rm for\ even} \ N.}}
Hence, for odd $N$ we can absorb the $\pm$ label in extending the range of $\theta$ to be in $[0,4\pi)$ (as in $SU(N)/\Z_N$).  However, for even $N$ the periodicity of $\theta $ is still $2\pi$ and the two classes of theories are not continuously connected, as in $SO(N)$ (indeed,  $SO(5)$  is equivalent to $Sp(2)/\Z_2$).  This fact is consistent with the discussion in \WittenNV, which shows the existence of half instantons for $Sp(N)/\Z_2$ with odd $N$, and the absence of such instantons for even $N$.

Before ending this sub-section we would like to comment on the $Sp(N)$ vs.\ $\widetilde{Sp}(N)$ theories of \WittenXY\ (see also \HananyFQ).  These two theories are obtained by a shift of $\theta$ by $\pi$.  Hence, even using the traditional classification of theories based on $\theta \in [0,2\pi)$ they are not new, and they should not be confused with the theories we discuss above.

\subsec{Pure $\CN=1$ SYM theories}

Next we examine the consequences of this discussion for the $\CN=1$ pure SYM theory.
The $Sp(N)$ gauge theory has a global $\Z_{2(N+1)}$ R-symmetry, and its instanton factor is $\eta = \Lambda^{3(N+1)}$.  The quantum theory has $N+1$ vacua with \IntriligatorNE
\eqn\lambdalambdaspn{\langle\lambda\lambda \rangle = \epsilon_{N+1}(2 \eta)^{1\over N+1},}
with $\epsilon_{N+1}$ an $N+1$'st root of unity.  In these vacua the $\Z_{2(N+1)}$ symmetry is spontaneously broken to $\Z_2$.  These vacua are related by shifting $\theta $ by $2\pi$.
In all of these vacua the Wilson lines in $(z_e,z_m)=(1,0)$ exhibit an area law, signaling confinement.

The $(Sp(N)/\Z_2)_\pm$ theories also have these $N+1$ vacua, and the vacua are still related by shifting $\theta$ by $2\pi$.  But the details depend on whether $N$ is even or odd.

For even $N$ the theory with $\theta$ is the same as the theory with $\theta+2\pi$, and hence all these vacua are related by the $\Z_{2(N+1)}$ symmetry.  The nontrivial line operators exhibit area law in all the vacua of the $(Sp(N)/\Z_2)_-$ theory, and they exhibit a perimeter law associated with an unbroken gauge $\Z_2$ symmetry in all the vacua of the $(Sp(N)/\Z_2)_+$ theory.

For odd $N$ the $N+1$ vacua are not all related by a symmetry.  The discrete R-symmetry of the $(Sp(N)/\Z_2)_\pm$ theories is only $\Z_{N+1}$.  This discrete symmetry is spontaneously broken to $\Z_2$, which is in the Lorentz group.  The $N+1$ vacua are split into two orbits of the discrete symmetry, with $(N+1)/2$ vacua in each.  In the vacua in one orbit the nontrivial line operators of $(Sp(N)/\Z_2)_+$ exhibit an area law, and those of $(Sp(N)/\Z_2)_-$ have a perimeter law associated with an unbroken $\Z_2$ gauge symmetry. The two behaviors are exchanged in the vacua of the other orbit.

Repeating the argument for $SU(N)/\Z_N$, this discussion suggests that the $Sp(N)$ theory on $\R^3 \times \S^1$ has $N+1$ vacua.  For even $N$ the $(Sp(N)/\Z_2)_-$ theory has $N+1$ vacua, while the $(Sp(N)/\Z_2)_+$ has $2(N+1)$ vacua. For odd $N$, both of the $(Sp(N)/\Z_2)_\pm$ theories have $3(N+1)/2$ vacua.

\subsec{S-duality in $\CN=4$ SYM}

\topinsert
\tabskip1em plus1fill\relax
{
\halign to\hsize{  \hfil #   &  \hfil # \hfil \cr
 & $\vcenter{\hbox{\epsffile{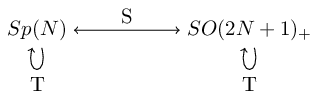}}}$ \cr
\noalign{\bigskip}
even $N$ & $\vcenter{\hbox{\epsffile{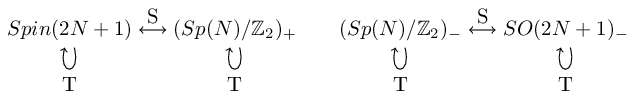}}}$ \cr
\noalign{\bigskip}
odd $N$ & $\vcenter{\hbox{\epsffile{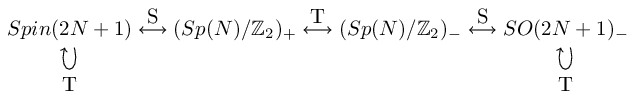}}}$ \cr
\noalign{\bigskip}
}}
\figcaption\sporbits{S-duality orbits of the $\CN=4$ SYM theory with $\g=so(2N+1), sp(N)$.}
\endinsert

The $\CN=4$ SYM theory has an S-duality transformation exchanging $\g=sp(N)$ and $\g^*=spin(2N+1)$. The
generator of the S-duality group in this case takes $\tau \to -1/(2\tau)$, and the mapping of the electric and magnetic weights also involves an extra factor of $2$; by an abuse of notation we will refer to this generator as $S$. The full duality group in this case is not $SL(2,\Z)$; it is generated by the $S$ generator and by $\tau \to \tau+1$. We already discussed how $\tau \to \tau+1$ permutes the different theories, and it is easy to determine how various choices of line operators are mapped to each other under the $S$ generator. Let us denote by $L\subset \Z_2\times \Z_2$ the set of charges of line operators of the original $sp(N)$ theory.
The set $L$ contains only one element in addition to $(z_e,z_m)=(0,0)$.
The $Sp(N)$ theory contains $(z_e,z_m)=(1,0)$. Thus, the dual theory contains lines with charges $(z_e,z_m)=(0,1)$, meaning that is an $SO(2N+1)_+$ theory.
The $(Sp(N)/\Z_2)_+$ theory contains $(z_e,z_m)=(0,1)$. Then, the dual theory contains lines with charges $(z_e,z_m)=(1,0)$, meaning that this is an $Spin(2N+1)$ theory.
Finally, the $(Sp(N)/\Z_2)_-$ theory contains lines with $(z_e,z_m)=(1,1)$.  Therefore, the dual theory contains lines with charges $(1,1)$, and this is an $SO(2N+1)_-$ theory.
The full transformations of these theories under S-duality are described in \sporbits. For even values of $N$ there are three separate orbits, each containing two theories. For odd values of $N$ there is one orbit containing the $Sp(N)$ and $SO(2N+1)_+$ theories, and the other four theories are in a second orbit.

\newsec{$\g = so(N)$ for even $N$}

In section 3 we already studied $Spin(N)$ and $SO(N)$. However, when $N$ is even
there are additional possibilities of gauge group $G$ for $\g=so(N)$.
The discussion depends on whether $N$ is divisible by $4$ or not.
Note that in these cases the magnetic dual algebra is also $\g=so(N)$.

\subsec{$N=2\, \mod\, 4$}

Let us begin with the simpler case when $N$ is not divisible by $4$, $N > 2$.
Then, the center of $\tilde G=Spin(N)$ is $\bC=\Z_4$. The two spinor representations transform as
$(\pm i)$ under the generator of the center.
The possible groups are $Spin(N)$, $Spin(N)/\Z_2=SO(N)$ or $Spin(N)/\Z_4$.
The first two cases were discussed already in section~3.

Let us consider then $G=Spin(N)/\Z_4=SO(N)/\Z_2$.
The class of charges of line operators is characterized by $(z_e,z_m)\in \Z_4\times \Z_4$.
The only purely electric lines have $(z_e,z_m)=(0,0)$. The maximality of the set of line operators
requires the existence of a line with charge $(z_e,z_m)=(n,1)$, where $n=0,1,2,3$.
Every choice is allowed by mutual locality.
So, we have the four theories $(Spin(N)/\Z_4)_{n\ \mod\ 4}$.
By the Witten effect, shifting $\theta$ by $2\pi$ sends $(z_e,z_m)$ to $(z_e\pm z_m,z_e)$,
where the sign is $+$ when $N\equiv -2\ \mod\ 8$ and $-$ when $N\equiv +2\ \mod\ 8$.\foot{%
The authors thank Ho Tat Lam for noticing a small error in the previous version of the paper.
The dependence of the sign on $N$ modulo 8 will be explained in the discussions around (6.13) below.
}
This shifts $n$ by $\pm1$, or in other words we have
\eqn\sotwomodfour{
(Spin(N)/\Z_4)_{n}^{{\theta+2\pi}}
=(Spin(N)/\Z_4)_{{n\pm 1\ \mod\ 4}}^{{\theta}}.
}
As a special case note that for $Spin(6)=SU(4)$ this agrees with our discussion of $SU(N)/\Z_N$ groups in section \sunzn.
The $\theta$-angle has periodicity $8\pi$, consistent with the existence of quarter-instantons in this case \WittenNV.

In the $\CN=1$ SYM theory with the $Spin(N)/\Z_4$ gauge groups, the global symmetry is $\Z_{(N-2)/2}$, and the $(N-2)$ vacua spontaneously break this to $\Z_2$. The vacua are split into $4$ classes. In each class the basic line operator of one of the $(Spin(N)/\Z_4)$ theories has a perimeter law, associated with a discrete $\Z_4$ gauge symmetry, the line operator of one of the other theories has a perimeter law associated with a discrete $\Z_2$ gauge symmetry, and those of the other two theories have an area law. Thus, when we compactify these theories on $\R^3\times \S^1$, they have $(N-2)+(N-2)/2+(N-2)/4+(N-2)/4=2(N-2)$ supersymmetric vacua.

\subsec{$N=0\, \mod\, 4$: centers and charges}

Next, let us come to the more interesting case when $N$ is a multiple of $4$. In this case there are two inequivalent spinor representations, that are not complex conjugates.
The center of $\tilde G=Spin(N)$ is now $\Z_2^S\times \Z_2^C$.
Here the generator of $\Z_2^S$ acts on one spinor representation by $(-1)$ and on the other representation by $(+1)$,
and the action of the generator of $\Z_2^C$ on the first spinor representation is by $(+1)$ and on the second by $(-1)$. The vector representation appears in the product of the two spinor representations, so it transforms as $(-1)$ under both $\Z_2$'s.
Let us denote the diagonal subgroup of  $\Z_2^S\times \Z_2^C$ as $\Z_2^V$.
We now have the following possibilities for the quotients:
$G=Spin(N)$, $G~=~Spin(N)/\Z_2^V=SO(N)$, $G=Spin(N)/\Z_2^S=Ss(N)$, $G=Spin(N)/Z_2^C=Sc(N)$,
and $G=Spin(N)/(\Z_2\times \Z_2)=SO(N)/\Z_2$.
 The $Sc(N)$ theory is related to the $Ss(N)$ theory by the $\Z_2$ outer-automorphism of $Spin(N)$, so it does not need to be treated separately. Nevertheless, the two theories have different spectra of line operators, and it will be useful to distinguish them when we discuss the action of S-duality later.
The group $Ss(N)$ is sometimes called the semispin group in the mathematical literature.
We note in passing that $Ss(32)$ is the gauge group of the Type I superstring in ten dimensions.

The weight lattice, both electric and magnetic, is given by the union of
$\Z^{N/2}$ and $(\Z^{N/2} + ({1\over 2},{1\over 2},\cdots,{1\over 2},{1\over 2}))$.
The weights are divided into four classes forming $\Z_2\times \Z_2$. The adjoint class $(0,0)$ contains elements of $\Z^{N/2}$ such that the sum of their coordinates is even, and the vector class $(1,1)$ contains the other elements of $\Z^{N/2}$ (in particular $\lambda_V=(1,0,\cdots,0)$). The two spinor classes $(1,0)$ and $(0,1)$ include the weight vectors
\eqn\spinorweight{
\lambda_S=(+{1\over 2},+{1\over 2},\cdots,+{1\over 2},+{1\over 2}),\qquad
\lambda_C=(+{1\over 2},+{1\over 2},\cdots,+{1\over 2},-{1\over 2}),
} respectively (and all other weight vectors given by these weights plus elements of the adjoint class).
Note that $\lambda_S\cdot \lambda_S=\lambda_C\cdot\lambda_C$ is even when $N=8d$, and is odd when $N=8d+4$, whereas $\lambda_S\cdot\lambda_C$ is odd when $N=8d$ and even when $N=8d+4$. This will lead to some differences between these two cases.

The classes of charges of line operators are now labeled by
\eqn\sofourcharge{
(z_{e,S},z_{e,C};z_{m,S},z_{m,C}) \in (\Z_2\times \Z_2)\times (\Z_2\times \Z_2).
}
The inner product determining the mutual locality condition is then
\eqn\localitysofour{\eqalign{
z_{e,S} z_{m,S}' - z_{m,S} z_{e,S}' +
z_{e,C} z_{m,C}' - z_{m,C} z_{e,C}' = 0\ \mod\ 2, &\quad  (N=8d+4) \cr
z_{e,S} z_{m,C}' - z_{m,C} z_{e,S}' +
z_{e,C} z_{m,S}' - z_{m,S} z_{e,C}' = 0\ \mod\ 2. &\quad (N=8d) \cr
}
}
For $N > 4$, shifting the $\theta$-angle by $2\pi$ modifies the charges according to
\eqn\shiftofchargesofour{
(z_{e,S},z_{e,C};z_{m,S},z_{m,C})
\to
(z_{e,S}+z_{m,S},z_{e,C}+z_{m,C};z_{m,S},z_{m,C}).
}
All the classes are periodic under $\theta \to \theta+4\pi$, consistent with the fact that these theories have half-instantons but do not have quarter-instantons \WittenNV.

\subsec{$N=0\, \mod\, 4$: Possible choices of lines}
\subseclab\sozeromodfour

After these preparations, it is straightforward to classify all possible sets of charges of line operators (we discuss here the cases of $N > 4$).
The cases $G=Spin(N)$ and $G=SO(N)$ have already been discussed in section~3.
We found $SO(N)_\pm$ there.

Let us next consider $G=Spin(N)/\Z_2^S=Ss(N)$. The purely electric lines have charges
in the classes $(0,0;0,0)$ and $(1,0;0,0)$.
The maximality of the set of charges requires that there are lines with charges
\eqn\magneticchargessofour{
\eqalign{
(0,n;0,1)  & \quad ({\rm if\ }N=8d+4), \cr
(0,n;1,0)  & \quad ({\rm if\ }N=8d). \cr
}
}
Here $n$ can be either $0$ or $1$, and correspondingly we have theories $Ss(N)_\pm$. The effect of the $\theta$-angle is easy to see:
\eqn\shiftss{
\eqalign{
Ss(N)^{{\theta+2\pi}}_{+}=Ss(N)^{{\theta}}_{-}  & \quad ({\rm if\ }N=8d+4),\cr
Ss(N)^{{\theta+2\pi}}_{\pm}=Ss(N)^{{\theta}}_{\pm}  & \quad ({\rm if\ }N=8d).\cr
}
}
Note that our discussion, including the last line, is consistent with \sotheta\ when $N=8$, as $Ss(8)$ and $SO(8)$ are then equivalent by an outer automorphism of $Spin(8)$.

Let us finally consider the case $G=Spin(N)/\Z_2\times \Z_2= SO(N)/\Z_2$.
The purely electric lines have charges $(0,0;0,0)$. The maximality of the set of charges then requires that there are lines with charges
\eqn\magneticmostinteresting{
(n_{SS},n_{SC};1,0),\quad
(n_{CS},n_{CC};0,1),
}
for some numbers $\{n_{SS},n_{SC},n_{CS},n_{CC}\}$ which can be $0$ or $1$.
The mutual locality imposes the constraint
\eqn\constraint{
\eqalign{
n_{SC}=n_{CS} & \quad ({\rm if\ }N=8d+4), \cr
n_{SS}=n_{CC} & \quad ({\rm if\ }N=8d), \cr
}
} and nothing else. In total, we find eight theories.
We label them as $(SO(N)/\Z_2)\ns{n_{SS}}{n_{SC}}{n_{CS}}{n_{CC}}$
with the understanding that the relation \constraint\ is satisfied.
Under the shift of the $\theta$-angle by $2\pi$, we have
\eqn\shiftA{
(SO(N)/\Z_2)\ns{n_{SS}}{n_{SC}}{n_{CS}}{n_{CC}}^{\theta+2\pi}  =
(SO(N)/\Z_2)\ns{n_{SS}+1}{,n_{SC}\phantom{+1}}{n_{CS}\phantom{+1}}{,n_{CC}+1}^{\theta}
}
and the eight theories fall into four orbits under the shift of $\theta$.

The generalization of our previous discussions of $\CN=1$ SYM to all of these cases is straightforward.

\subsec{$\CN=4$}

The S-duality of $\CN=4$ SYM maps the gauge algebra $so(N)$ to itself when $N$ is even,
but it permutes the global structures and the choice of lines.

\topinsert
\tabskip1em plus1fill\relax
{
\halign to\hsize{  \hfil #   &  \hfil # \hfil \cr
 & $\vcenter{\hbox{\epsffile{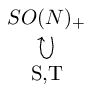}}}$ \cr
\noalign{\bigskip}
odd $d$: & $\vcenter{\hbox{\epsffile{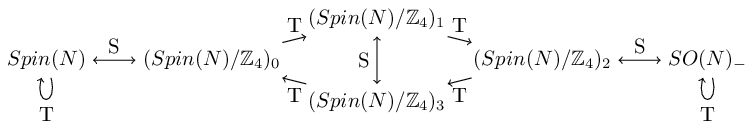}}}$ \cr
\noalign{\bigskip}
even $d$: & $\vcenter{\hbox{\epsffile{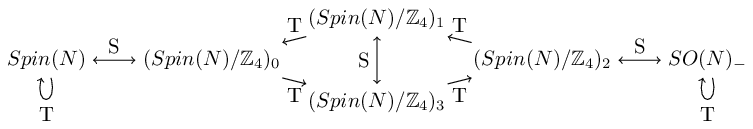}}}$ \cr
\noalign{\bigskip}
}}
\figcaption\sosix{S-duality orbits of the $\CN=4$ SYM theories with $\g=so(4d+2)$.}
\endinsert

When $N=4d+2$, the structure is exactly as in $su(4)\simeq so(6)$.
Therefore, under the S-generator we have the mapping:
\eqn\Sdualfourkplustwo{
\matrix{
Spin(N) & \longleftrightarrow & (Spin(N)/\Z_4)_0 \cr
SO(N)_- & \longleftrightarrow & (Spin(N)/\Z_4)_2 \cr
(Spin(N)/\Z_4)_1 & \longleftrightarrow & (Spin(N)/\Z_4)_3 \cr
}}
while $SO(N)_+$ is left invariant. Including the shift of theta discussed above, this implies that (as for $su(4)$) six of the theories are in one orbit of $SL(2,\Z)$, while the $SO(N)_+$ theory is invariant under the full $SL(2,\Z)$ duality group.
The orbits are shown in \sosix.\foot{%
The authors thank Ho Tat Lam for noticing a small error in the previous version of the paper, where we mistakenly used the same duality orbits for even $d$ and odd $d$.
}

\pageinsert

\bigskip
\centerline{\epsffile{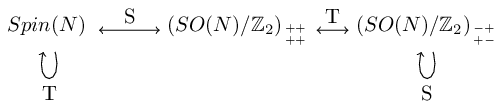}}
\bigskip
\centerline{\epsffile{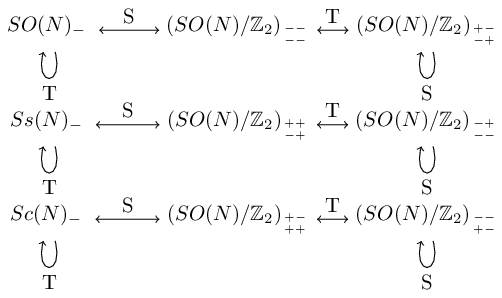}}
\centerline{\epsffile{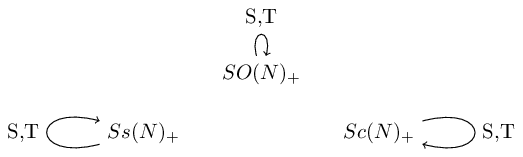}}
\figcaption\soeight{S-duality orbits of the $\CN=4$ SYM theories with $\g=so(8d)$.}
\bigskip

\centerline{\epsffile{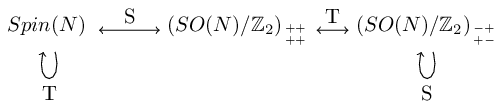}}
\bigskip
\centerline{\epsffile{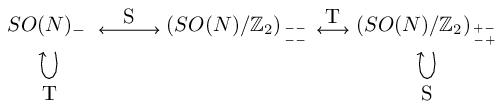}}
\bigskip
\centerline{\epsffile{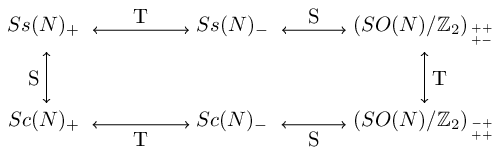}}
\bigskip
\centerline{\epsffile{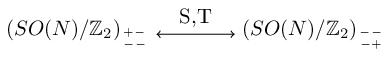}}
\bigskip
\centerline{\epsffile{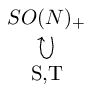}}
\figcaption\sotwelve{S-duality orbits of the $\CN=4$ SYM theories with $\g=so(8d+4)$.}
\endinsert

When $N=4d$, the $15$ possible theories are generally permuted by the full S-duality group.
For $N=8d$, three theories are invariant under the full $SL(2,\Z)$, which are the $SO(N)_+$, $Ss(N)_+$ and $Sc(N)_+$ theories. The other $12$ theories fall into $4$ separate orbits, each containing $3$ different theories. For example, the $Spin(N)$ theory is mapped by the $S$ generator to the $(SO(N)/\Z_2)\ns0000$ theory, which is mapped by the $T$ generator to the $(SO(N)/\Z_2)\ns1001$ theory, which is mapped to itself by the $S$ generator.
The orbits are given in \soeight.
For $N=8d+4$, the only theory which is fully invariant under $SL(2,\Z)$ is the $SO(N)_+$ theory. The other theories are divided into two orbits containing $3$ theories (one of which contains $Spin(N)$, and the other contains $SO(N)_-$), one orbit of size $6$, and another orbit of size $2$ (containing the $(SO(N)/\Z_2)\ns0111$ and $(SO(N)/\Z_2)\ns1110$ theories, which map to each other under $T$, and to themselves under $S$).
The orbits are shown in \sotwelve.\foot{%
Previous versions had a couples of typos in the duality orbits of $SO(N)_-$ in \soeight\ (corrected in v2) and \sotwelve\ (corrected in v5).
The authors thank Oren Bergman and Sebastian Rauch for corrections.
}

\newsec{Summing over bundles in the Euclidean path integral}

When the spatial slice of the spacetime of our gauge theories is a nontrivial $3$-manifold, the states in the Hilbert space are characterized by discrete electric and magnetic charges related to the center of the covering group (see \WittenNV\ and references therein). As we will discuss below, different charges are allowed in different theories of the types we discussed above, with different Hilbert spaces.
Correspondingly, the Euclidean partition functions of these different theories on general 4-manifolds are different(though they are the same on $\S^4$ and $\S^3\times
\S^1$).  In this last section, we discuss how the Euclidean partition function distinguishes the different theories (with different sets of line operators), using the standard $\theta$-angle $\theta$, and in some cases its discrete generalizations. We begin in section 6.1 with an analysis of the Euclidean path integral and
the possible $\theta$-angles that appear in it. In section 6.2 we show how to map these $\theta$-angles to
the theories we discussed in the previous sections. In section 6.3 we generalize to the case where the center contains more than one factor. In section 6.4 we discuss surface operators, and their relation to the $\Z_k$ gauge theories that we encountered above.

\subsec{$\theta$-angles in the Euclidean path integral}

The standard $\theta$-angle $\theta$ we are familiar with is a phase $i\theta \ell$ in the action,
where $\ell$ is the instanton number. We always use the physicists' normalization of the instanton number, where a small instanton (that exists on any space) has instanton number 1.
For a  theory with simply-connected gauge group $\tilde G$, the instanton number is integer on any manifold, and $\theta$ has $2\pi$ periodicity.
When the gauge group $G$ is not simply connected, the instanton number can be fractional
on a nontrivial manifold, or in the presence of a line operator. Then, the periodicity of the $\theta$-angle is  $2\pi x$, where $x$ is an integer determined by $G$. This $x$ can further depend on whether we allow non-spin manifolds or not;
in this paper we assume for simplicity that every four-dimensional manifold we deal with is spin.

When the gauge group $G$ is not simply connected, the gauge bundle has additional characteristic classes, other than the instanton number. Let us begin with the example of $G=Spin(N)/\Z_2=SO(N)$. In this case we have the Stiefel-Whitney classes $w_{2,4}$ of the gauge bundle, which are cohomology classes of degree $2$ and $4$ defined in $\Z_2$, in addition to the integer-valued Pontryagin class $p_1$ (related to the instanton number) which is of degree 4. It is known (see \Thomas\ and references therein) that these classes are related by
\eqn\wvsp{
	p_1 = \CP(w_2)+2w_4 \quad \hbox{mod}\ 4.
} Here, on the right-hand side, we use the Pontryagin square operation $\CP$, which sends a degree-$2d$ mod-$2$ class to a degree-$4d$ mod-$4$ class:
\eqn\pontryaginsquare{
	H^{2d}(X,\Z_2)\ni v \mapsto \CP(v) \in H^{4d}(X,\Z_{4}).
}
This is a generalization of the usual square $v^2$, in the sense that
\eqn\pontvsusualsquare{
	\CP(v)=v^2 \ \mod \ 2.
}
Furthermore, by $2w_4$, we mean the image of $w_4$ under the homomorphism
$2: H^{4}(X,\Z_2)\to H^4(X,\Z_4)$
coming from the homomorphism $\Z_2\to \Z_4$ sending $1$ mod $2$ to $2$ mod $4$.\foot{%
This is not the standard notation in mathematics, but can be justified e.g.~by considering $w_4$ as a cochain in $C^2(X,\Z_2)$, lifting it to $C^2(X,\Z)$, multiplying it by $2$ there, and reducing it to $C^2(X,\Z_4)$.
The notation $\CP(w_2)/2$ below is not standard in mathematics either, for which 
we also provided a definition in this version.
The authors thank Greg Moore for a discussion on these points.
}
In the following, we write the integral of $\CP(w_2)$ over the manifold simply as $\CP(w_2)$, to lighten the notation.

For $G=SO(N)$ with $N\ge 4$,  $p_1=2\ell$, where $\ell$ is the instanton number. 
On a spin manifold the intersection product is even.  
Therefore $\CP(w_2)$ is $0$ or $2$ mod $4$.
We then define $\CP(w_2)/2$ to be $0$ or $1$ mod $2$, respectively.
The equation \wvsp\ then says that $\ell$ can only be integer, and that $\CP(w_2)/2$ is even or odd if $(w_4-\ell)$ is even or odd.
Therefore, the $\theta$-angle $\theta$ is defined modulo $2\pi$, and $\CP(w_2)/2$ (or, equivalently, $w_4$) gives rise to an independent discrete $\theta$-angle; we can weigh the configurations in our Euclidean path integral by a phase
\eqn\sotheta{
i\theta\ell  + in\pi{\CP(w_2)\over 2}, \quad \theta\sim \theta+2\pi, \ n=0,1.
}

For $G=SO(3)$, $w_4=0$ and $p_1=4\ell$, where $\ell$ is the instanton number. Again, $\CP(w_2)$ is defined mod $4$ and is even. Then, \wvsp\ says that $\ell$ can be half-integer, and $\CP(w_2)/2$ is even or odd according to whether $\ell$ is integer or half-integer.
Therefore, the $\theta$-angle $\theta$ is defined modulo $4\pi$, and $\CP(w_2)/2$ does not give an independent discrete $\theta$-angle. The term we can write in the action is then
\eqn\sothreetheta{
i\theta\ell , \quad \theta\sim \theta+4\pi~.}
Note that we can still equivalently use \sotheta\ instead.

In principle we can study possible types of discrete $\theta$-angles by having a look at all the degree-4 characteristic classes of non-simply-connected groups $G$.
More mathematically, this corresponds to the classification of $H^4(BG,U(1))$, where $BG$ is the classifying space and $H^4$ refers to the standard singular cohomology. (For a readable account on $BG$, see e.g.~\DijkgraafPZ.) Instantons that can be put on an open set of $\R^4$ are measured by the homotopy group $\pi_3(G)=\pi_3(\tilde G)$, which gives a natural subgroup of $H^4(BG,U(1))$.
When $H^4(BG,U(1))$ is $\simeq \Z$, the usual $\theta$-angle with enlarged periodicity covers all the possible choices of the phase in the Lagrangian.
When this group has more structure, we need additional discrete $\theta$-angles to fully describe possible phases in the Lagrangian.
Unfortunately we could not find a comprehensive discussion of $H^4(BG,U(1))$ in the mathematics literature, although scattered results on many $G$'s can be found.

Instead, we can proceed by following \refs{\DoldWhitney,\WittenNV}.
Let $\tilde G$ be a simple simply-connected group. Then its quotient by a subgroup of the center is either $\tilde G/\Z_k$ or $\tilde G/\Z_2\times \Z_2$; the latter only occurs when $\tilde G=Spin(4d)$.
Fix a closed spin four-manifold $X$. Consider a $G$-bundle on it. It cannot always be lifted to a $\tilde  G$-bundle. This obstruction is controlled by
\eqn\generalsw{
w_2 \in H^2(X,\Z_k)
}
when $G=\tilde G/\Z_k$ and by
\eqn\moregeneralsw{
w_2^{(1)},\ w_2^{(2)} \in H^2(X,\Z_2)
}
when $G=\tilde G/\Z_2\times \Z_2$.\foot{%
These obstruction classes, controlling whether a $G$ bundle can be lifted to a $\tilde G$ bundle and taking values in $H^2(X,\Z_k)$, are often called  (generalized) Stiefel-Whitney classes in physics literature.
This usage is not standard in mathematics, where the Stiefel-Whitney classes instead refer to characteristic classes $w_k\in H^k(X,\Z_2)$ associated to a real bundle.
There, $w_2$ is related to the lifting of an $SO$ bundle to a $Spin$ bundle 
but other $w_k$'s are not.
The degree-2 class controlling the lifting from $PSU(N)$ to $U(N)$ or $SU(N)$ is sometimes called the Brauer class in mathematical literature, 
but unfortunately there seems to be no standard mathematical terminology for the class controlling the lifting from $G$ to $\tilde G$.
The authors thank Dan Freed and Michael Hopkins for the correspondences on these issues.
}
Let us consider two $G$-bundles $E$, $E'$ with the same $w_2$.
Then their difference can be measured by an {\it integer} instanton number in $\pi_3(G)\simeq \pi_3(\tilde G)$ (see e.g.~footnote 14 of \WittenNV).
In other words, the instanton number mod 1 is uniquely determined by $w_2$, and the phase including the contributions from various characteristic classes can always be written as
\eqn\generaltheta{
i\theta\ell  + i{2n\pi \over k}{\CP(w_2)\over 2}, \quad \theta\sim \theta+2\pi, \ n=0,1,\ldots,k-1
} when $G=\tilde G/\Z_k$, and
\eqn\spintheta{
i\theta\ell  + i \sum_j n_j\pi{\CP(w_2^{(j)})\over 2} + in_{12} \pi {w_2^{(1)}\cdot w_2^{(2)}},
\quad \theta\sim \theta+2\pi, \quad n_{1}, n_{12}, n_2=0,1
} when $G=\tilde G/\Z_2\times \Z_2$.
In order for \generaltheta\ to be well-defined, we need a squaring operation $\CP(w_2)$ such that
$\CP(w_2)/2$ is well-defined modulo $k$. For even values of $k$ we use here
the general Pontryagin square operation \BrowderThomas, which sends
\eqn\generalpontryaginsquare{
H^{2d}(X,\Z_{k})  \ni v\to \CP(v)\in H^{4d}(X,\Z_{2 k}).
}
$\CP(w_2)/2$ is then defined  in $\Z_k$, since our manifold $X$ is spin.
For odd values of $k$, by an abuse of notation we also denote by $\CP(v)$
the standard cup product
\eqn\cupsquare{
H^{2d}(X,\Z_{k})  \ni v\to \CP(v)=v^2\in H^{4d}(X,\Z_{k}).
}
Here, since $k$ is odd, $2$ is invertible modulo $k$, and therefore $\CP(v)/2$ is again defined modulo $k$.
As in the $SO(N)$ case, we expect that adding the characteristic class $w_4$ to \generaltheta\ should not add any
additional information; it would be interesting to verify this by a careful analysis of the relevant bundles.

In some cases, as in the $SO(3)$ theory we discussed above, the term proportional to $\CP(w_2)/2$ can be replaced by a change in the periodicity of $\theta$.
This can be found by referring to the computation in \WittenNV\ of the instanton number mod 1 in terms of $w_2$, for the case where $G=\tilde G/\bC$ where $\bC$ is the full center.
For reference we list the results here. When $\tilde G\neq Spin(4d)$, the fractional part of the instanton number is equal to
\eqn\fracpart{\ell = s {\CP(w_2)\over 2} \quad\hbox{mod}\ 1,}
where $s$ is given by\foot{%
The authors thank Ho Tat Lam for noticing a small error in the previous version of the paper, where $s$ was stated to be $+1/4$ for both $N=8d+2$ and $N=8d+6$. 
For more details, see the footnote in Sec.~2.3.5 of \CordovaUOB.
}
\eqn\shift{
\vcenter{\halign{
\hfil $#$ &  \quad $#$ \hfil \quad  &  \quad $#$ \quad \hfil & \quad $#$ \hfil \cr
\tilde G & & \bC & s  \cr
SU(N) & & \Z_N & 1/N   \cr
Spin(N) &   (N=2d+1) & \Z_2 & 0  \cr
Spin(N)  &  (N=8d+2) & \Z_4 & -1/4  \cr
Spin(N)  &  (N=8d+6) & \Z_4 & +1/4  \cr
Sp(N) & (N=2d) & \Z_2& 0  \cr
Sp(N) & (N=2d+1) & \Z_2 & 1/2 \cr
E_6 &  & \Z_3 & 2/3 \cr
E_7 &  & \Z_2 & 1/2 \cr
}}
} Here $Spin(N)$ is understood to have $N\ge 5$.
The analogous statement for the group $G=Spin(N)/\Z_2\times \Z_2$ is
\eqn\shiftspineightn{
\ell = \cases{{1\over 2}w_2^{(1)}\cdot w_2^{(2)}\quad\hbox{mod}\ 1 & for $N=8d$\cr
{1\over 2}\left(\CP(w_2^{(1)})/2 +  \CP(w_2^{(2)})/2\right)\quad\hbox{mod}\ 1
 & for $N=8d+4$ ~.}}
For example, $Sp(N)/\Z_2$ has the following properties that can be verified using \Borel:
\item{$N $ even:} The periodicity of the $\theta$-angle is $2\pi$, $\CP(w_2)$ is linearly independent of the instanton number, and the difference between the instanton number and $\CP(w_2)/2$ should be an independent mod-2 degree-4 characteristic class $w_4$.
\item{$N$ odd:} The periodicity of the $\theta$-angle is $4\pi$, $\CP(w_2)/2$ is determined by the instanton number, and there should not be an independent mod-2 degree-4 characteristic class $w_4$.

\noindent Similarly, the periodicity of the $\theta$-angle for $SU(3)/\Z_3$ and for $E_7/\Z_2$ is $6\pi$ and $4\pi$, respectively, and there should not be an independent $w_4$ mod $3$ or mod $2$, respectively. This is also true \refs{\KMS,\KM}.

It is straightforward to generalize this analysis to arbitrary semi-simple groups.
The universal covering group $\tilde G=\prod_j \tilde G_j$ is the product of simple simply-connected groups,
and the gauge group $G$ can be written as $G=\tilde G/\prod_i \Z_{k_i}$.
For each $\Z_{k_i}$, we have a Stiefel-Whitney class $w_2^{(i)}$ defined mod $k_i$.
Then, the most general form of the $\theta$-angles is
\eqn\verygeneraltheta{
i\sum_j \theta_j \ell_j + i\sum_i {2n_i\pi \over k_i}{\CP(w_2{}^{(i)})\over 2} + i\sum_{i_1<i_2} {2n_{i_1,i_2}\pi \over \gcd(k_{i_1},k_{i_2})} w_2^{(i_1)}\cdot w_2^{(i_2)}
}  where $\theta_j\sim \theta_j+2\pi$, $n_i=0,\ldots,k_i-1$, $n_{i_1,i_2}=0,\ldots,\gcd(k_{i_1},k_{i_2})-1$, and
the cup product $w_2^{({i_1})}\cdot w_2^{({i_2})}$ is taken after reducing both $w_2^{({i_1},{i_2})}$ modulo $\gcd(k_{i_1},k_{i_2})$.

\subsec{$\theta$-angles and line operators}

We want to understand how the values of the discrete $\theta$-angles that we found above are related to the different choices of line operators, which we described in the previous sections.
The analysis is essentially a generalization of the discussion in \WittenNV\ of discrete electric and magnetic charges associated with the center of the gauge group, incorporating our discrete $\theta$-angles.
 For simplicity we discuss the case $G=\tilde G/\Z_k$. The generalization to $G=Spin(4n)/\Z_2\times\Z_2$ and to general semi-simple groups is straightforward.

Consider the insertion of a straight line operator in $\R^4$. The topology of spacetime in the presence
of this line operator is $\S^2\times (0,\infty)_r \times \R_t$.
Here $\S^2$ is the sphere surrounding the line, $r$ is the radial direction with appropriate boundary conditions at $r\to 0,\infty$, and $t$ is the direction along the line, which we view as the Euclidean time direction.
The discrete magnetic charge carried by this line operator is given by
\eqn\magnetic{
m=\int_{\S^2} w_2 \in \Z_k,
}
where $m$ is precisely the magnetic charge $z_m$ of the line operator that we defined before.

The discrete electric charge of this configuration should equal to the electric charge $z_e$ of the line operator. As defined in \WittenNV, it is given by the eigenvalue
of a gauge transformation
\eqn\electricgaugetr{
g: (0,\infty)_r \to G,  \quad g(r=0)=g(r=\infty)=1
} which belongs to the homotopy class of the generator of $\pi_1(G)=\Z_k$.
In the path integral formalism, this is measured by the phase assigned to a gauge configuration on
\eqn\euclideanconfig{
\S^2\times (0,\infty)_r \times [0,\beta]_t =: \S^2 \times \T^2,
 }
where the configurations at $t=0$ and $t=\beta$ are identified by the gauge transformation specified by $\hat m$. This is defined by having a $G$-bundle with
\eqn\electric{
\hat m=\int_{\T^2} w_2 \in \Z_k.
}
We then have $w_2=m[\S^2]+{\hat m}[\T^2]$ and
\eqn\wtwosquared{
\int_{\S^2\times \T^2} {\CP(w_2)\over 2}= m\hat m \in \Z_k.
}
With the generalized $\theta$-angles \generaltheta\ in the action, this gives the phase
\eqn\phasegiven{2\pi i{n \over k}  m\hat m.}
This means that  the discrete electric charge of the setup is $nm\in \Z_k$.
Thus, the theory with a discrete theta parameter $n$ includes line operators with magnetic charge $z_m=m$ and electric charge $z_e=n m {\rm \ mod\ } k$.

For concreteness let us set $G=SU(N)/\Z_k$, $\tilde G=SU(N)$ and $kk'=N$. Then the analysis so far says that the line operators with minimal magnetic charge have the charge $(z_e,z_m)=(n,k')\in \Z_N\times \Z_N$,
reproducing the charge lattice $L_{n}$ in \sunno\ when $k=N$, or $L_{k,n}$ in \sunnok\ in the general case. Therefore we see that the subscript $n$ in $(SU(N)/\Z_k)_n$ is indeed the coefficient $n$ in the path integral phase \generaltheta.

We can also repeat the same discussion in the Hamiltonian formalism of \WittenNV, putting the theory on
some spatial 3-manifold $Y$. In this formalism
every two-cycle is associated with a magnetic charge $m_i$ in $\Z_k$, associated with $w_2$, and every
one-cycle is associated with an electric charge $e_i$ in $\Z_k$, defined through a gauge transformation
as above. The Hilbert space is divided into sectors labeled by the discrete charges of all the one-cycles and two-cycles. We claim that in a theory with given line operators, the Hilbert space includes those sectors such that for every pair of a one-cycle and dual two-cycle, their charges $(e_i, m_i)$ must be among the charges of the allowed line operators\foot{We do not discuss manifolds with torsion cycles here.}. This is because the insertion of a topologically trivial line operator with charges $(z_e,z_m)$ creates around it charges $(e,m)$ equal to those of the line operator as discussed above, while the insertion of a line operator wrapped around a one-cycle shifts the associated $(e_i,m_i)$ by $(z_e,z_m)$.
As an example, the theory with gauge group ${\tilde G}$ contains only the sectors $(e_i,m_i=0)$, while
the theory with gauge group $G$ and index $0$ contains only the sectors with $\Z_k$ charges $(e_i=0,m_i)$.

In \WittenNV, the Witten index of the pure ${\cal N}=1$ SYM theory was computed for all simple gauge groups and for each sector $(e_i,m_i)$. If we want to count, say, the number of supersymmetric vacua of the pure ${\cal N}=1$ SYM theory on $\R^3\times \S^1$ for one of the theories that we described in the previous sections, we need to sum the Witten index over the charges $(e,m)$ (associated to the $\S^1$ and to the dual 2-cycle) that are in the list of charges of allowed line operators. This precisely agrees with the counting of vacua that we derived in the previous sections.

To relate to our Euclidean discussion, we consider the same theory on $Y\times \S^1$. Every one-cycle becomes a two-cycle when including the $\S^1$, and the magnetic charge ${\hat m}_i$ on that 2-cycle is precisely the discrete Fourier transform of its electric charge $e_i$. The phase coming from the $\CP(w_2)$ term in the path
integral is exactly $2\pi n \sum_i m_i {\hat m}_i / k$, which is equivalent to saying that in the sector with magnetic charge $m_i$ we project onto the electric charge $e_i=n m_i$. As we described above, in
some cases we can replace the parameter $n$ by extending the range of the standard $\theta$-angle, but
in general it is not possible to do this.

\subsec{Lattice of charges of line operators and $\theta$-angles: the general case}

To study the effect of the phases in the path integral in the general case $G=\tilde G/\prod_i \Z_{k_i}$, including the case $G=Spin(4d)/(\Z_2 \times \Z_2)$, it is better to first study the most general charge lattice.

Let $\tilde G=\prod_r \tilde G_r$, where $\tilde G_r$ are simple and simply-connected groups.
Let $\bC_r$ be the center of $\tilde G_r$. When $\tilde G_r$ is $Spin(4d)$, we need to decompose $\bC_r=\Z_2\times \Z_2$. We write
\eqn\totalcenter{
\bC=\prod_r \bC_r=\prod_{s} \Z_{a_{s}}.
}
Take a subgroup
\eqn\subgroup{
\bH=\prod_i \Z_{k_i}\subset \bC =\prod_{s} \Z_{a_{s}},
}
and consider a theory with gauge group $G=\tilde G/\bH$. Note that $\Z_{k_i}$ here
can be various nontrivial subgroups of combinations of $\Z_{a_s}$ in \totalcenter.
The charge of a line operator belongs to a class
\eqn\generalclass{
(z_e,z_m)\in \bC\times \bC,
}where we have
\eqn\chargescharges{
z_e=(n_1\,\mod\,a_1,n_2\,\mod\,a_2,\ldots),\quad
z_m=(m_1\,\mod\,a_1,m_2\,\mod\,a_2,\ldots).
}
We choose the basis of the magnetic charges so that the inner product determining the mutual locality is
\eqn\innerproduct{
\sum_{s} {n_{s} m'_{s} - m_{s} n'_{s}\over a_s} = 0\ \mod\ 1.
}
In view of \localitysofour, this means that we take, for example,
\eqn\mappingofcharges{
\eqalign{
(n_1,n_2;m_1,m_2)=(z_{e,S},z_{e,C};z_{m,S},z_{m,C}) & \quad (\tilde G=Spin(8d+4)), \cr
(n_1,n_2;m_1,m_2)=(z_{e,S},z_{e,C};z_{m,C},z_{m,S}) & \quad (\tilde G=Spin(8d)). \cr
}
}

Purely electric lines have charges $z_e\in \Gamma\subset \bC$, where $\Gamma$ is the set of classes where $\prod_i \Z_{k_i}$ acts trivially.
The mutual locality and the maximality of the set of the charges then require that $z_m$ in \generalclass\ is in $\prod_i\Z_{k_i}\subset \bC$.
Denote the generator of $\Z_{k_i}$ by $g_i\in \bC$.  For each $i$ there is a line with magnetic charge $g_i$ and we need to choose its electric charge $\nu_i\in \bC$.  Since we can multiply by purely electric line operators, we can classify the allowed charges $\nu_i$ modulo elements of $\Gamma$.  These can be represented by an element of $\bC/\Gamma=\prod_j \Z_{k_j}$, which we denote by $\nu_{ij}$ modulo $k_j$. Recall, $i$ labels the line and $j$ labels which $\Z_{k_j}$ subgroup of the $\bH$ we are considering.  These numbers completely specify the allowed charges.
Next, mutual locality requires that
\eqn\consequence{
{\nu_{ij} \over k_j} = {\nu_{ji} \over k_i}\ \mod\ 1.
} The general solution to this is given by
\eqn\solution{
{\nu_{ij}\over k_j} = {\nu_{ji}\over k_i}={n_{ij} \over \gcd(k_j,k_i)}, \qquad n_{ij}=0,1,\ldots,\gcd(k_j,k_i)-1.
}

Therefore, the most general choice of the set of charges of line operators is specified by numbers
\eqn\specification{
n_i\equiv \nu_{ii} =0,1,\ldots, k_i-1,\qquad
n_{ij} =0,1,\ldots, \gcd(k_i,k_j)-1.
}
They nicely match the parameters in the most general phase in the path integral \verygeneraltheta. Repeating the analysis in section~6.2, it is easy to see that indeed the phase \verygeneraltheta\ reproduces the charges of the dyonic line operators described above.

Let us verify that for $G=Spin(N)/\Z_2\times \Z_2$ (when $N$ is divisible by 4) this reproduces our previous statements in section \sozeromodfour. The discrete $\theta$-angles are given in \spintheta, by $n_{11},n_{12}=n_{21},n_{22}$, all taking the values $0$ or $1$.
Using \mappingofcharges\ and \innerproduct, we see that we have line operators with charges
\eqn\AAA{
\eqalign{
(n_{11},n_{12};1,0), \quad (n_{21},n_{11};0,1) &\quad (N=8d+4)\,, \cr
(n_{12},n_{11};1,0), \quad (n_{22},n_{21};0,1) &\quad (N=8d)\,,
}
} which indeed satisfy the constraint \constraint.
Also, the modulo-1 equality of the instanton number and the combination of the Stiefel-Whitney classes $w_2^{(1)}$, $w_2^{(2)}$ in \shiftspineightn\ correctly translate to the shift of the parameters $n_{SS}$, $n_{CS}$, $n_{SC}$ and $n_{CC}$ in \shiftA.

\subsec{Surface operators}

In our gauge theories we can also define various surface operators. The ones that are most interesting
for us are the ones associated with the center of the gauge group.

Let us consider surface operators in a gauge group $G$ with a nontrivial center $\bC$, for example $SU(N)$ with its center $\Z_N$.  We can define
a surface operator associated with every element $g \in \bC$. An insertion of this surface
operator on some surface $M$ is defined by requiring that when we go around the surface and come back to the same point, the gauge
transformation parameter is multiplied by $g$.  This is nontrivial, since the gauge transformation involved is not periodic in $G$, but all the gauge invariant operators (local operators and line operators) are still well-defined in the presence of this surface operator. Every Wilson line is associated with a representation of $G$ and hence an element in the center $g'\in \bC$.  The expectation value of a Wilson line with $g'$ and a surface operator with $g$ then depends on their linking number $q$ through a phase.  For $\bC=\Z_k$ it is $\exp(2\pi i g g' q/ k)$.

Next, let us consider the surface operators when the gauge group $G$ has a nontrivial first homotopy group $\pi_1(G)$,  for example $G=SU(N)/\Z_N$.
We can construct here a surface operator that is S-dual to the one of the previous
paragraph (see section 4.3 of \GukovSN). In this case there is a Stiefel-Whitney class $w_2$ associated with bundles of this group, which gives an element of the center when integrated over surfaces, and we can define an insertion of a surface operator on $M$ to modify the path integral by a phase; when $\pi_1(G)=\Z_k$ this phase is $\exp(2\pi i \int_M w_2 / k)$. As in the previous paragraph, the expectation values of 't~Hooft lines carrying center charges and of these surface operators depends on their linking number through a phase (which is in $\Z_k$ for $\pi_1(G)=\Z_k$).

We saw in the previous sections that in some cases we get at low energies discrete $\Z_k$ gauge theories. Such gauge theories have line operators and surface operators, labeled by an element of $\Z_k$. They obey a similar relation to the one described in the previous two paragraphs; the correlation function of a line operator and a surface operator has a $\Z_k$ phase given by the product of their charges and of the linking number of the line with the surface. One way to realize such a discrete gauge theory is to have a 1-form potential $A$ and a 2-form potential $B$, with standard quantization of their fluxes (such that their integrals over closed cycles are integer multiples of $2\pi$), and with an action \refs{\MaldacenaSS, \BanksZN}
\eqn\topaction{S = {k\over 2\pi i} \int B \wedge dA.}
In this realization, the $\Z_k$ line operators are Wilson lines of $A$, and the $\Z_k$ surface operators are given by  exponentials of integrals of $B$. When we flow from any of our gauge theories to a $\Z_k$ gauge theory, the basic line operator of that gauge theory becomes the line operator of this $\Z_k$ theory, and the surface operators described in the previous two paragraphs become the surface operators of this $\Z_k$ theory.

We can use this discussion also to relate the different theories discussed above. For instance, suppose we have an $SU(N)/\Z_N$ theory and we want to turn it into an $SU(N)$ theory. We can couple the $SU(N)/\Z_N$ theory to a $\Z_N$ gauge theory (realized by $A$ and $B$ fields as above), and add to the action a term
\eqn\newterm{\int d^4x B \wedge w_2.}
The field $B$ then serves as a Lagrange multiplier that sets the discrete magnetic charge to zero\foot{Note that the equation of motion of $A$ implies that $B$ is flat, and that as in other places we ignore possible issues related to torsion.}. The magnetic line operators of the $SU(N)/\Z_N$ theory are no longer allowed since they are not gauge-invariant under gauge transformations of $B$, and the surface operators of the $SU(N)/\Z_N$ theory become trivial. On the other hand, we introduce new line and surface operators from the $\Z_N$ gauge theory, which behave like line and surface operators of $SU(N)$. In a similar way, one can go from an $SU(N)$ theory to an $SU(N)/\Z_N$ theory, by coupling a $\Z_N$ gauge theory such that the $B$ field of the $\Z_N$ theory makes the surface operators of the $SU(N)$ theory trivial. By similar manipulations we can
relate all the theories we discussed above (that have the same Lie algebra $\g$), by coupling them to an appropriate discrete gauge theory.

\bigskip

\noindent {\bf Acknowledgments}

We would like to thank T.~Dumitrescu, G.~Festuccia, D.~Freed, D.~Gaiotto, S.~Giacomelli, J.~Gomis, K.~Intriligator,  T.~Okuda, S.~Razamat, M.~Strassler, S.~Sugimoto, B.~Willett, and E.~Witten for many useful discussions.  We would like to thank G.~W.~Moore in particular, for helpful discussions and also for his detailed and thoughtful comments on the draft.
We would also like to thank Oren Bergman for pointing out an important typo in a previous version of this paper.
OA is the Samuel Sebba Professorial Chair of Pure and Applied Physics, and he is supported in part by a grant from the Rosa and Emilio Segre Research Award, by an Israel Science Foundation center for excellence grant, by the German-Israeli Foundation (GIF) for Scientific Research and Development, and by the Minerva foundation with funding from the Federal German Ministry for Education and Research. OA gratefully acknowledges support from an IBM Einstein Fellowship at the Institute for Advanced Study.  The work of NS was supported in part by DOE grant DE-FG02-90ER40542 and by the United States-Israel Binational Science Foundation (BSF) under grant number~2010/629.  YT's research is supported in part by MEXT/JSPS grant KAKENHI-25870159, and in part by the WPI research initiative through IPMU, University of Tokyo.
Any opinions, findings, and conclusions or recommendations expressed in this
material are those of the authors and do not necessarily reflect the views of the funding agencies.

\def\listrefs{
\immediate\closeout\rfile\writestoppt
\bigskip\baselineskip=\footskip{\noindent {\bf References}\hfill}\medskip{\parindent=20pt%
\frenchspacing\escapechar=` \input \jobname.refs\vfill\eject}\nonfrenchspacing}

\listrefs

\end